# Alternative N=2 Supergravity in Singular Five Dimensions with Matter/Gauge Couplings[1]

Hitoshi NISHINO[2]

*Department of Physics*
*University of Maryland*
*College Park, MD 20742-4111*

and

Subhash RAJPOOT[3]

*Department of Physics & Astronomy*
*California State University*
*Long Beach, CA 90840*

**Abstract**

We present an extended study of our previous work on an alternative five-dimensional $N=2$ supergravity theory that has a single antisymmetric tensor and a dilaton as a part of supergravity multiplet. The new fields are natural Neveu-Schwarz massless fields in superstring theory. Our total matter multiplets include $n$ copies of vector multiplets forming the sigma-model coset space $SO(n,1)/SO(n)$, and $n'$ copies of hypermultiplets forming the quaternionic Kähler manifold $Sp(n',1)/Sp(n') \times Sp(1)$. We complete the couplings of matter multiplets to supergravity with the gauged group of the type $SO(2) \times Sp(n') \times Sp(1) \times H \times [U(1)]^{n-p+1}$ for an arbitrary gauge group $H$ with $p \equiv \dim H + 1$, and the isotropy group $Sp(n') \times Sp(1)$ of the coset $Sp(n',1)/Sp(n') \times Sp(1)$ formed by the hypermultiplets. We also describe the generalization to singular 5D space-time as in the conventional formulation

---

[1]This work is supported in part by NSF grant # PHY-98-02551.
[2]E-Mail: nishino@nscpmail.physics.umd.edu
[3]E-Mail: rajpoot@csulb.edu

# 1. Introduction

The importance of supergravity in 5D space-time manifests itself in many contexts, such as the supersymmetrization [1][2] of Randall-Sundrum type brane-world scenario [3], namely, gauged supergravity in singular 5D space-time. In ref. [2], the introduction of a 4th-rank antisymmetric tensor $A_{\mu\nu\rho\sigma}$ made it easier to handle supergravity in such a singular space-time with the orbifold-type singularity $S^1/\mathbb{Z}_2$. Another important aspect of 5D supergravity is related to what is called holographic anti-de-Sitter and superconformal field theory (AdS/SCF) correspondence, namely the conjecture that the large $N$ limit of $SU(N)$ superconformal field theories in 4D are dual equivalent to supergravity on AdS space-time in 5D [4][5]. In both of these aspects of 5D supergravity, the presence of the 5D cosmological constant, via the gauging of the $N=2$ automorphism group $SL(2,\mathbb{R}) = Sp(1)$ (or its $SO(2)$ subgroup) plays a crucial role.

The conventional on-shell formulation of $N=2$ supergravity in 5D was initiated in [6] in which an arbitrary number of vector multiplets is coupled to supergravity, and generalized further in [7][8][9]. However, in these formulations [6][7][8][9], the dilaton field as one of the important NS fields does not have manifest dilaton scale invariance. Moreover, an additional complication is that the tensor fields in [6][7][8][9] appear in symplectic pairs, obeying the 'self-duality' condition in odd space-time dimensions, and therefore the single antisymmetric tensor field $B_{\mu\nu}$ as another important NS field [10] is mixed up with other tensor fields. In order to overcome these drawbacks in these on-shell formulations [6][7][8][9], we may try an off-shell formulation as an alternative, but such a formulation lacks the manifest $\sigma$-model geometry formed by scalars, which is 'hidden' at the off-shell level before eliminating auxiliary fields. This is similar to the 4D case of Kähler manifold structure in on-shell $N=1$ supergravity [11] which is hidden in the off-shell formulation.

In our previous paper [12], we have proposed an alternative on-shell $N=2$ supergravity multiplet in 5D, which has an irreducible field content larger than the conventional one [6][8][9] including an antisymmetric tensor and a dilaton fields that are Neveu-Schwarz (NS) massless fields in superstring theory [10]. Our supergravity multiplet has the field content $(e_\mu{}^m, \psi_\mu{}^A, B_{\mu\nu}, \chi^A, A_\mu, \sigma)$ with 12+12 on-shell degrees of freedom, where the fünfbein $e_\mu{}^m$, the gravitini $\psi_\mu{}^A$, and the graviphoton $A_\mu$ are the same as the conventional $N=2$ supergravity [6][7][9][8], while an antisymmetric tensor $B_{\mu\nu}$, a dilatino $\chi^A$, and a dilaton $\sigma$ are our new field content. Among these, the antisymmetric tensor $B_{\mu\nu}$ and the dilaton $\sigma$ are natural NS massless fields in superstring theory [10].

In the present paper, we will continue the study of our alternative on-shell $N=2$ supergravity [12] coupled to $n$ copies of vector multiplets and $n'$ copies of hypermultiplets. In our formulation, the dilaton and the antisymmetric fields as the important NS fields are treated separately from other scalars. Our $n$ scalars $\varphi^\alpha$ form the coordinates of the $\sigma$-model coset $SO(n,1)/SO(n)$, while the $4n'$ scalars $\phi^{\underline{\alpha}}$ form the coordinates of the $\sigma$-model coset $Sp(n',1)/Sp(n') \times Sp(1)$ [12]. We present the general gaugings of our system in the presence of hypermultiplets that were not given in our previous paper [12], and consider our supergravity in singular 5D space-time, as the supersymmetrization of Randall-Sundrum brane-world scenario [3], following the prescription of [2] for dealing with the orbifold-type $S^1/\mathbb{Z}_2$ singularity.

This paper is organized as follows: In section 2, we review our $N = 2$ alternative supergravity as a notational preparation before gaugings. In section 3, we give the general treatment for the gauging of an arbitrary non-Abelian gauge group that has nothing to do with the coset $Sp(n',1)/Sp(n') \times Sp(1)$. Section 4 is devoted to our main focus in the present paper, namely, to show how to gauge the automorphism group $Sp(1) = SL(2,\mathbb{R})$ of $N = 2$ supersymmetry, or more generally the whole isotropy group $Sp(n') \times Sp(1)$ of the coset $Sp(n',1)/Sp(n') \times Sp(1)$, in the presence of hypermultiplets, which was not accomplished in our previous paper [12]. As a by-product, we will give the most general case of gauging of the total group $SO(2) \times Sp(n') \times Sp(1) \times H \times [U(1)]^{n-p+1}$ for an arbitrary gauge group $H$ with $p \equiv \dim H + 1$, and the isotropy group $Sp(n') \times Sp(1)$ of the coset $Sp(n',1)/Sp(n') \times Sp(1)$ formed by the hypermultiplets. Section 5 is for the formulation of our alternative supergravity in a singular 5D space-time, with the orbifold-type singularity $S^1/\mathbb{Z}_2$, i.e., the supersymmetrization [1][2] of Randall-Sundrum brane-world scenario [3]. Section 6 is for our conclusion, while the important notations and conventions are given in the Appendix.

## 2. Coupling of Vector Multiplets and Hypermultiplets to 5D, $N = 2$ Supergravity

We start with reviewing the couplings of 5D, $N = 2$ supergravity to vector multiplets and hypermultiplets [12], before general non-Abelian gaugings. The field content of the multiplet of supergravity is $(e_\mu{}^m, \psi_\mu{}^A, A_\mu{}^I, B_{\mu\nu}, \chi^A, \varphi^\alpha, \lambda^{aA}, \phi^{\underline{\alpha}}, \psi^{\underline{a}})$ with $12+12$ on-shell degrees of freedom [12]. Here $\mu, \nu, \cdots$ are for the curved world indices, while $m, n, \cdots$ are local Lorentz with the metric $(\eta_{mn}) = \text{diag.}(-,+,+,+,+)$, $e_\mu{}^m$ is the fünfbein, $\psi_\mu{}^A$ is the gravitino with $A = 1, 2$ for **2**-representation of the automorphism group $Sp(1) = SL(2,\mathbb{R})$ for the $N = 2$ supersymmetry. The raising/lowering of the indices $A, B, \cdots$ is performed by the $Sp(1)$ metric $\epsilon_{AB}, \epsilon^{AB}$, and therefore special attention is needed for superscript/subscript of these indices, in particular, their inner products. As in [12], we use here $Sp(1) = SL(2,\mathbb{R})$ notation instead of $SU(2)$ as the automorphism group, in order to make all the bosonic fields manifestly real, just for simplicity. The vectors $A_\mu{}^I$ ($I = 0, 1, 2, \cdots, n$) form the $(\mathbf{n+1})$-representation of $SO(n,1)$ in the coset $SO(n,1)/SO(n)$ [13][14]. The $\varphi^\alpha$ ($\alpha = 1, 2, \cdots, n$) are the $n$-dimensional $\sigma$-model coordinates of the coset $SO(n,1)/SO(n)$, $\lambda^{aA}$ ($a = 1, 2, \cdots, n$) are in the **n**-representation of $SO(n)$, $\phi^{\underline{\alpha}}$ ($\underline{\alpha} = 1, 2, \cdots, 4n'$) are the $4n'$-dimensional coordinates of the quaternionic Kähler manifold $Sp(n',1)/Sp(n') \times Sp(1)$, and $\psi^{\underline{a}}$ ($\underline{a} = 1, 2, \cdots, 2n'$) are in the **2n'**-representation of $Sp(n')$. As described in [12], this is the combination of our multiplet of supergravity $(e_\mu{}^m, \psi_\mu{}^A, A_\mu, B_{\mu\nu}, \chi^A, \sigma)$, $n$ copies of the vector multiplets $(C_\mu, \lambda^A, \varphi)$, and $4n'$ copies of the hypermultiplets $(\phi^{\underline{\alpha}}, \psi^{\underline{a}})$. In particular, the graviphoton $A_\mu$ is identified with the zero-th component $A_\mu{}^0$, while the $n$ copies of the vector field $C_\mu$ from the vector multiplets renamed as $A_\mu{}^1, A_\mu{}^2, \cdots, A_\mu{}^n$, combined into the unified notation $A_\mu{}^I$ ($I = 0, 1, 2, \cdots, n$). Since the indices $I, J, \cdots$ are with the indefinite metric $(\eta_{IJ}) = \text{diag.}(-,+,+,\cdots,+)$, we make the raising/lowering of these indices explicit. Note that our multiplet of supergravity is distinct from the conventional one $(e_\mu{}^m, \psi_\mu{}^A, A_\mu)$ [6], in which only the fünfbein, gravitino and the graviphoton form the irreducible field content.

The geometrical relationships associated with the coset $SO(n,1)/SO(n)$ are conveniently listed up as [13][12]

$$[H_{ab}, H_{cd}] = \delta_{bc}H_{ad} - \delta_{ac}H_{bd} + \delta_{ad}H_{bc} - \delta_{bd}H_{ac} ~, \tag{2.1a}$$

$$[H_{ab}, K_c] = \delta_{bc}K_a - \delta_{ac}K_b ~, \quad [K_a, K_b] = +2H_{ab} ~. \tag{2.1b}$$

$$L_A{}^I \partial_\alpha L_I{}^B = \tfrac{1}{2}A_\alpha{}^{ab}(H_{ab})_A{}^B + V_\alpha{}^a(K_a)_A{}^B ~, \quad A_{\alpha b}{}^c = L_b{}^I \partial_\alpha L_I{}^c ~, \tag{2.1c}$$

$$(H_{ab})_c{}^d = \delta_{ac}\delta_b{}^d - \delta_{bc}\delta_a{}^d ~, \quad (K_a)_{b(0)} = (K_a)_{(0)b} = -\sqrt{2}\delta_{ab} ~, \tag{2.1d}$$

$$L_I{}^A L_A{}^J = \delta_I{}^J ~, \quad L_A{}^I L_I{}^B = \delta_A{}^B ~, \tag{2.1e}$$

$$L_I \equiv L_I{}^{(0)} ~, \quad L^I \equiv L_{(0)}{}^I ~, \quad L_I L^I = +1 ~, \quad L_a{}^I L_I \equiv 0 ~, \quad L_I{}^a L^I \equiv 0 ~, \tag{2.1f}$$

$$L_{IJ} \equiv \eta_{AB} L_I{}^A L_J{}^B = -L_I L_J + L_I{}^a L_{Ja} \tag{2.1g}$$

$$L_{IJ} L^J = -L_I ~, \quad L_{IJ} L_a{}^J = +L_{Ia} ~, \tag{2.1h}$$

$$D_\alpha L_I = \partial_\alpha L_I = -\sqrt{2} L_I{}^a V_{\alpha a} ~, \quad D_\alpha L_I{}^a = -\sqrt{2} L_I V_\alpha{}^a ~, \quad \partial_\alpha L_{IJ} = 0 ~, \tag{2.1i}$$

$$[D_\alpha, D_\beta] L_I{}^a = -2(V_\alpha{}^a V_\beta{}^b - V_\beta{}^a V_\alpha{}^b) L_{Ib} ~, \tag{2.1j}$$

$$R_{\alpha\beta}{}^{ab} = -2(V_\alpha{}^a V_\beta{}^b - V_\beta{}^a V_\alpha{}^b) ~, \quad R = -2n(n-1) \leq 0 ~, \tag{2.1k}$$

$$D_\alpha X_a \equiv \partial_\alpha X_a + A_{\alpha a}{}^b X_b ~, \tag{2.1\ell}$$

which are self-explanatory exactly in the same notation as in [12]. The Cartan decomposition of the $SO(n,1)$ Lie algebra is dictated by the $SO(n)$ generators $H_{ab}$ and the coset generators $K_a$, satisfying (2.1). The indices $a, b, \cdots = (1), (2), \cdots, (n)$ are for the vectorial representation of $SO(n)$. The indices $A, B, \cdots = ((0),a), ((0),b), \cdots = (0), (1), (2), \cdots, (n)$ are for the local coordinates on $Sp(n',1)/Sp(n') \times Sp(1)$.[4] In other words, $A, B, \cdots = ((0),a), ((0),b), \cdots$ are

---

[4]The indices $A, B, \cdots$ used both for the **2**-representations and for these local Lorentz coordinates are

the $(n+1)$-dimensional extension of the original $n$-dimensional indices $a$, $b$, $\cdots$. The indices $I$, $J$, $\cdots = 0, 1, \cdots, n$ are for the curved coordinates, while $\alpha$, $\beta$, $\cdots = 1, 2, \cdots, n$ are for the coordinates on $Sp(n',1)/Sp(n') \times Sp(1)$. The raising/lowering of the indices $A$, $B$, $\cdots$ is performed by the metric tensor $(\eta_{AB}) = \text{diag.}(-,+,+,\cdots,+)$. The Maurer-Cartan form made of $L_I{}^A$ decomposes as in (2.1c). Eq. (2.1d) gives the explicit components of $H$'s and $K$'s, while (2.1e) - (2.1i) are relevant orthonormality relations. Eq. (2.1i) - (2.1$\ell$) are for the $SO(n)$ covariant derivative $D_\alpha$.

As for the geometry related to the quaternionic Kähler manifold $Sp(n',1)/Sp(n') \times Sp(1)$ we start with the representative $L^{\underline{a}\alpha}$, which satisfies the Maurer-Cartan form for the coset $Sp(n',1)/Sp(n') \times Sp(1)$ [15][16][9][7][8][12]:

$$L^{-1}\partial_{\underline{\alpha}} L = A_{\underline{\alpha}}{}^i T^i + A_{\underline{\alpha}}{}^{\underline{I}} T_{\underline{I}} + V_{\underline{\alpha}}{}^{\underline{a}A} K_{\underline{a}A} \ , \tag{2.2a}$$

$$g_{\underline{\alpha}\underline{\beta}} V_{\underline{a}A}{}^{\underline{\alpha}} V_{\underline{b}B}{}^{\underline{\beta}} = \epsilon_{\underline{ab}} \epsilon_{AB} \ , \qquad V_{\underline{\alpha}\underline{a}A} V_{\underline{\beta}}{}^{\underline{a}B} = +\tfrac{1}{2} g_{\underline{\alpha}\underline{\beta}} \delta_A{}^B - \tfrac{1}{2} F_{\underline{\alpha}\underline{\beta}}{}^i (T^i)_A{}^B \ , \tag{2.2b}$$

$$J_{\underline{\alpha}\underline{\beta}}{}^i = -(T^i)_A{}^B (V_{\underline{\alpha}\underline{a}B} V_{\underline{\beta}}{}^{\underline{a}A} - V_{\underline{\beta}\underline{a}B} V_{\underline{\alpha}}{}^{\underline{a}A}) = -\tfrac{1}{2} F_{\underline{\alpha}\underline{\beta}}{}^i \ , \tag{2.2c}$$

where $T_{\underline{I}}$ ($\underline{I} = 1, 2, \cdots, n'(n'+1)/2$) are the generators of $Sp(n')$, $T^i$ ($i = 1, 2, 3$) are the generators of $Sp(1)$, while $K_{\underline{a}A}$ are the coset generators of $Sp(n',1)/Sp(n') \times Sp(1)$ [15][16]. All of these equations involve the vielbein $V_{\underline{\alpha}}{}^{\underline{a}A}$ for this quaternionic Kähler manifold.

With all other details of geometry skipped, our lagrangian before gaugings is [12]

$$\begin{aligned}
e^{-1}\mathcal{L}_0 = & -\tfrac{1}{4}R - \tfrac{1}{2}(\overline{\psi}_\mu \gamma^{\mu\nu\rho} D_\nu \psi_\rho) - \tfrac{1}{12} e^{-4\sigma} G_{\mu\nu\rho}^2 - \tfrac{1}{4} e^{-2\sigma}(L_I{}^a L_{Ja} + L_I L_J) F_{\mu\nu}{}^I F^{\mu\nu J} \\
& - \tfrac{1}{2}(\overline{\lambda}^a \gamma^\mu D_\mu \lambda_a) - \tfrac{1}{2} g_{\alpha\beta}(\partial_\mu \varphi^\alpha)(\partial^\mu \varphi^\beta) - \tfrac{3}{4}(\partial_\mu \sigma)^2 - \tfrac{1}{2}(\overline{\chi} \gamma^\mu D_\mu \chi) \\
& - \tfrac{1}{2} g_{\underline{\alpha}\underline{\beta}}(\partial_\mu \phi^{\underline{\alpha}})(\partial^\mu \phi^{\underline{\beta}}) - \tfrac{1}{2}(\overline{\psi}^{\underline{a}} \gamma^\mu D_\mu \psi_{\underline{a}}) \\
& + \tfrac{i}{\sqrt{2}} V_\alpha{}^a (\overline{\psi}_\mu \gamma^\nu \gamma^\mu \lambda_a) \partial_\nu \varphi^\alpha + \tfrac{\sqrt{3}i}{2}(\overline{\psi}_\mu \gamma^\nu \gamma^\mu \chi) \partial_\nu \sigma + i V_{\underline{\alpha}}{}^{\underline{a}A} (\overline{\psi}_\mu \gamma^\nu \gamma^\mu \psi_{\underline{a}}) \partial_\nu \phi^{\underline{\alpha}} \\
& - \tfrac{i}{4\sqrt{2}} e^{-\sigma}(\overline{\psi}_\mu \gamma^{\mu\nu\rho\sigma} \psi_\nu + 2\overline{\psi}^\rho \psi^\sigma) L_I F_{\rho\sigma}{}^I + \tfrac{i}{6\sqrt{3}} e^{-2\sigma}(\overline{\psi}_\mu \gamma^{\rho\sigma\tau} \gamma^\mu \chi) G_{\rho\sigma\tau} \\
& - \tfrac{1}{24} e^{-2\sigma}(\overline{\psi}_\mu \gamma^{\mu\nu\rho\sigma\tau} \psi_\nu - 6\overline{\psi}^\rho \gamma^\sigma \psi^\tau) G_{\rho\sigma\tau} - \tfrac{5}{72} e^{-2\sigma}(\overline{\chi} \gamma^{\mu\nu\rho} \chi) G_{\mu\nu\rho} \\
& - \tfrac{1}{2\sqrt{2}} e^{-\sigma}(\overline{\psi}_\mu \gamma^{\rho\sigma} \gamma^\mu \lambda_a) L_I{}^a F_{\rho\sigma}{}^I - \tfrac{1}{2\sqrt{6}} e^{-\sigma}(\overline{\psi}_\mu \gamma^{\rho\sigma} \gamma^\mu \chi) L_I F_{\rho\sigma}{}^I - \tfrac{i}{12\sqrt{2}} e^{-\sigma}(\overline{\chi} \gamma^{\mu\nu} \chi) L_I F_{\mu\nu}{}^I \\
& + \tfrac{i}{4\sqrt{2}} e^{-\sigma}(\overline{\lambda}^a \gamma^{\mu\nu} \lambda_a) L_I F_{\mu\nu}{}^I + \tfrac{1}{24} e^{-2\sigma}(\overline{\lambda}^a \gamma^{\mu\nu\rho} \lambda_a) G_{\mu\nu\rho} + \tfrac{1}{24} e^{-2\sigma}(\overline{\psi}^{\underline{a}} \gamma^{\mu\nu\rho} \psi_{\underline{a}}) G_{\mu\nu\rho} \\
& + \tfrac{i}{\sqrt{6}} e^{-2\sigma}(\overline{\chi} \gamma^{\mu\nu} \lambda_a) L_I{}^a F_{\mu\nu}{}^I - \tfrac{1}{4\sqrt{2}} e^{-\sigma}(\overline{\psi}^{\underline{a}} \gamma^{\mu\nu} \psi_{\underline{a}}) L_I F_{\mu\nu}{}^I \ , \tag{2.3}
\end{aligned}$$

yielding an invariant action $S_0$ under supersymmetry

$$\delta_Q e_\mu{}^m = +(\overline{\epsilon} \gamma^m \psi_\mu) \ , \qquad \delta_Q \sigma = +\tfrac{i}{\sqrt{3}}(\overline{\epsilon} \chi) \ ,$$

$$\delta_Q \psi_\mu{}^A = +D_\mu \epsilon^A + \tfrac{i}{6\sqrt{2}} e^{-\sigma}(\gamma_\mu{}^{\rho\sigma} - 4\delta_\mu{}^\rho \gamma^\sigma)\epsilon^A L_I F_{\rho\sigma}{}^I + \tfrac{1}{18} e^{-2\sigma}(\gamma_\mu{}^{\rho\sigma\tau} - \tfrac{3}{2}\delta_\mu{}^\rho \gamma^{\sigma\tau})\epsilon^A G_{\rho\sigma\tau} \ ,$$

*not* to be confused each other, as long as we keep track of the context they are used.

$$\delta_Q A_\mu{}^I = -\tfrac{i}{\sqrt{2}} e^\sigma L^I (\bar\epsilon \psi_\mu) + \tfrac{1}{\sqrt{6}} e^\sigma L^I (\bar\epsilon \gamma_\mu \chi) + \tfrac{1}{\sqrt{2}} e^\sigma (\bar\epsilon \gamma_\mu \lambda^a) L_a{}^I ~~,$$

$$\delta_Q B_{\mu\nu} = +e^{2\sigma} (\bar\epsilon \gamma_{[\mu} \psi_{\nu]}) + \tfrac{i}{\sqrt{3}} e^{2\sigma} (\bar\epsilon \gamma_{\mu\nu} \chi) - 2 L_{IJ} A_{[\mu|}{}^I (\delta_Q A_{|\nu]}{}^J) ~~,$$

$$\delta_Q \chi^A = -\tfrac{1}{2\sqrt{6}} e^{-\sigma} \gamma^{\mu\nu} \epsilon^A L_I F_{\mu\nu}{}^I + \tfrac{i}{6\sqrt{3}} e^{-2\sigma} \gamma^{\mu\nu\rho} \epsilon^A G_{\mu\nu\rho} - \tfrac{\sqrt{3}i}{2} \gamma^\mu \epsilon^A \partial_\mu \sigma ~~,$$

$$\delta_Q \varphi^\alpha = +\tfrac{i}{\sqrt{2}} V_a{}^\alpha (\bar\epsilon \lambda^a) ~~,$$

$$\delta_Q \lambda^{aA} = -\tfrac{1}{2\sqrt{2}} e^{-\sigma} \gamma^{\mu\nu} \epsilon^A L_I{}^a F_{\mu\nu}{}^I - \tfrac{i}{\sqrt{2}} \gamma^\mu \epsilon^A V_\alpha{}^a \partial_\mu \varphi^\alpha ~~,$$

$$\delta_Q \phi^{\underline{\alpha}} = +i V_{\underline{a}A}{}^{\underline{\alpha}} (\bar\epsilon^A \psi^{\underline{a}}) ~~,$$

$$\delta_Q \psi^{\underline{a}} = -i V_{\underline{\alpha}}{}^{\underline{a}A} \gamma^\mu \epsilon_A \partial_\mu \phi^{\underline{\alpha}} ~~, \tag{2.4}$$

up to quartic fermion (or quadratic fermion) terms in the lagrangian (or transformation rules). Here we have omitted the $Sp(1)$ indices $A$, $B$, $\cdots$ in the $Sp(1)$-invariant products, e.g., $(\bar\epsilon \gamma^m \psi_\mu) \equiv (\bar\epsilon^A \gamma^m \psi_{\mu A})$.

As in the usual dilaton couplings in supergravity [17], the antisymmetric field $B_{\mu\nu}$ and the vectors $A_\mu{}^I$ are scaled, when the dilaton $\sigma$ is shifted by a constant value:

$$\sigma \to \sigma + c ~~, \quad B_{\mu\nu} \to e^{2c} B_{\mu\nu} ~~, \quad A_\mu{}^I \to e^c A_\mu{}^I ~~, \tag{2.5}$$

where $c$ is an arbitrary constant parameter. This global symmetry controls the various exponential couplings of $\sigma$ in the lagrangian (2.3).

The various covariant derivatives and the field strength $G_{\mu\nu\rho}$ in these equations are given by

$$D_\mu \epsilon^A \equiv D_\mu(\omega) \epsilon^A + (\partial_\mu \phi^{\underline{\alpha}}) A_{\underline{\alpha}}{}^i (T^i \epsilon)^A ~~,$$

$$D_{[\mu} \psi_{\nu]}{}^A \equiv D_{[\mu|}(\omega) \psi_{|\nu]}{}^A + (\partial_{[\mu|} \phi^{\underline{\alpha}}) A_{\underline{\alpha}}{}^i (T^i \psi_{|\nu]})^A ~~,$$

$$D_\mu \chi^A \equiv D_\mu(\omega) \chi^A + (\partial_\mu \phi^{\underline{\alpha}}) A_{\underline{\alpha}}{}^i (T^i \chi)^A ~~,$$

$$D_\mu \lambda^{aA} \equiv D_\mu(\omega) \lambda^{aA} + (\partial_\mu \varphi^\alpha) A_\alpha{}^{ab} \lambda_b{}^A + (\partial_\mu \phi^{\underline{\alpha}}) A_{\underline{\alpha}}{}^i (T^i \lambda^a)^A ~~,$$

$$D_\mu \psi^{\underline{a}} \equiv D_\mu(\omega) \psi^{\underline{a}} + (\partial_\mu \phi^{\underline{\alpha}}) A_{\underline{\alpha}}{}^{\underline{I}} (T^{\underline{I}} \psi)^{\underline{a}} ~~,$$

$$G_{\mu\nu\rho} \equiv 3 \partial_{[\mu} B_{\nu\rho]} - 3 L_{IJ} F_{[\mu\nu}{}^I A_{\rho]}{}^J ~~, \tag{2.6}$$

where $A_\alpha{}^{ab}$ is the composite $SO(n)$ connection on the coset $SO(n,1)/SO(n)$, while $A_{\underline{\alpha}}{}^{\underline{I}}$ and $A_{\underline{\alpha}}{}^i$ are respectively the composite connections of $Sp(n')$ and $Sp(1)$ in $Sp(n',1)/Sp(n') \times Sp(1)$. The action of the generators $T^i$ or $T^{\underline{I}}$ is such as $(T^i \epsilon)^A \equiv (T^i)^{AB} \epsilon_B$ or $(T^{\underline{I}} \psi)^{\underline{a}} \equiv (T^{\underline{I}})^{\underline{ab}} \psi_{\underline{b}}$. Since we are not concerned with the quadratic fermionic terms in the transformation rule (2.4), the Lorentz connection $\omega_\mu{}^{rs}$ contains the usual unholonomy coefficients just made of the fünfbeins.

Compared with the conventional formulations [6], there is a similarity as well as basic difference. The similarity is that our tensor field $B_{\mu\nu}$ can be dualized into a vector field $B_\mu$ by a duality transformation so that the final field content will be $(e_\mu{}^m, \psi_\mu{}^A, A_\mu, B_\mu, \chi^A, \sigma)$. From this viewpoint, our system (2.1) is 'dual equivalent' to the conventional formulation with only one vector multiplet, in particular the dilaton field plays the coordinate of $SO(1,1)$, as usual in superstring theory. However, the caveat at this stage is that even though such a duality transformation is possible even after coupling vector multiplets, the resulting $\sigma$-model structure is qualitatively different from that given in the conventional formulations [6][7][9][8], as has been also explained in our previous paper [12].

As has been also stressed in [12], the antisymmetric field $B_{\mu\nu}$ and the dilaton $\sigma$ are the natural NS massless fields in superstring [10] or M-theory [18]. Therefore, it is more natural to have a supergravity with these fields in the point field theory limit. Another advantage of introducing an antisymmetric tensor $B_{\mu\nu}$ is associated with the recent development of non-commutative geometry in which the tensor $B_{\mu\nu}$ develops certain non-trivial constant value. We stress the fact that our supergravity multiplet contains the NS fields $B_{\mu\nu}$ and $\sigma$ as irreducible component fields, indicating that our supergravity is a more natural point field theory limit of superstring theory [10] or M-theory [18] than the conventional one [6][7][9][8].

## 3. Non-Abelian Gauging of Subgroup of $SO(n,1)$

We next establish general non-Abelian gaugings in the presence of vector multiplets and hypermultiplets. In this section, we consider the case that the gauged non-Abelian group $G$ has nothing to do with isotropy groups $Sp(n') \times Sp(1)$ in the coset $Sp(n',1)/Sp(n')$, but is just any other independent Lie group, which may be needed for more practical model building. Since all the $n$ copies of vectors in the vector multiplets together with the graviphoton in the multiplet of supergravity form the $(\mathbf{n+1})$-representation of $SO(n,1)$ in the coset $SO(n,1)/SO(n)$, we need special care for such non-Abelian gaugings. Such non-Abelian gauging has been performed in the conventional formulation [6], as well as recent works in [7][9][8], and in other dimensions such as in 7D [19]. In the formulation below, we will mainly follow the notation in [19], in which the coset space formed by the scalars in the vector multiplets is $SO(n,3)/SO(n) \times SO(3)$. This is slightly different from our coset $SO(n,1)/SO(n)$, but we still can take advantage of the similarity between them.

First of all, the non-Abelian gauge group $G$ should be the subgroup of $SO(n,1)$, and at the same time $\dim G = n+1$ should be satisfied, due to the coset structure to be maintained. Second, the structure constant $f_{IJ}{}^K$ should satisfy the relationship [19]

$$f_{IJK} \equiv f_{IJ}{}^L L_{LK} = f_{[IJK]} \quad . \tag{3.1}$$

where $L_{IJ}$ is the indefinite metric on $SO(n,1)/SO(n)$ as in section 2. This condition is satisfied when this indefinite metric $L_{IJ}$ is identified with the Cartan-Killing metric $\eta_{IJ}$ of $G$. To be specific, we can have $G = SO(2) \times H \times [U(1)]^{n-p+1}$, so that $\dim H = p-1$, $\dim G = n+1$, and

$$(L_{IJ}) \equiv (\eta_{IJ}) = \text{diag.}(-1, \eta_{I'J'}, +1, +1, \cdots, +1) ~, \tag{3.2}$$

where $I', J' = 1, 2, \cdots, p-1$ with $1 \le p \le n+1$. Also in (3.2), the first $-1$ is the Cartan-Killing metric of $SO(2)$ for the 0-th direction in the $(n+1)$-dimensions, $\eta_{I'J'}$ is that of $H$, while the last $(+1, +1, \cdots, +1)$ are the metrics for the Abelian factor groups $[U(1)]^{n-p+1}$. In the special case of $p = n+1$, there is no $U(1)$ factor group. This situation is similar to that in [19].

For such a gauge group $G$, we introduce the minimal coupling with the coupling constant $\overline{g}$. Typically, we have [19]

$$P_{\mu a} \equiv L_a{}^I \partial_\mu L = -\sqrt{2} V_{\alpha a} \partial_\mu \varphi^\alpha$$
$$\longrightarrow \quad \mathcal{P}_{\mu a} \equiv L_a{}^I (\partial_\mu L_I + \overline{g} f_{IJ}{}^K A_\mu{}^J L_K) \equiv L_a{}^I \mathcal{D}_\mu L_I \equiv -\sqrt{2} V_{\alpha a} \mathcal{D}_\mu \varphi^\alpha ~, \tag{3.3a}$$

$$A_{\mu a}{}^b \equiv (\partial_\mu \varphi^\alpha) A_{\alpha a}{}^b$$
$$\longrightarrow \quad \mathcal{A}_{\mu a}{}^b \equiv A_{\mu a}{}^b + \overline{g} f_{IJ}{}^K A_\mu{}^I L_a{}^J L_K{}^b ~, \tag{3.3b}$$

$$D_\mu \lambda^{aA} \equiv D_\mu(\omega) \lambda^{aA} + (\partial_\mu \varphi^\alpha) A_\alpha{}^{ab} \lambda_b{}^A + (\partial_\mu \phi^{\underline{\alpha}}) A_{\underline{\alpha}}{}^i (T^i \lambda^a)^A$$
$$\longrightarrow \quad \mathcal{D}_\mu \lambda^{aA} \equiv D_\mu(\omega) \lambda^{aA} + \mathcal{A}_\mu{}^{ab} \lambda_b{}^A + (\partial_\mu \phi^{\underline{\alpha}}) A_{\underline{\alpha}}{}^i (T^i \lambda^a)^A ~, \tag{3.3c}$$

$$G_{\mu\nu\rho} \quad \longrightarrow \quad \mathcal{G}_{\mu\nu\rho} \equiv 3 \left( \partial_{[\mu} B_{\nu\rho]} - L_{IJ} F_{[\mu\nu}{}^I A_{\rho]}{}^J + \tfrac{1}{3} \overline{g} f_{IJ}{}^K A_\mu{}^I A_\nu{}^J A_{\rho K} \right) ~. \tag{3.3d}$$

Eq. (3.3a) is none other than the standard minimal non-Abelian coupling for the adjoint index $I$. Needless to say, the structure constants $f_{IJ}{}^K$ with the indices $I, J, K$ for any of the $U(1)$ factor groups or $SO(2)$ are supposed to vanish. So effectively, only the indices $I', J', K'$ on $f_{IJ}{}^K$ remain. If we rewrite (3.3a) as

$$\mathcal{P}_\mu{}^\alpha \equiv V_a{}^\alpha \mathcal{P}_\mu{}^a \equiv -\sqrt{2} (\partial_\mu \varphi^\alpha - \overline{g} A_\mu{}^I \xi^\alpha{}_I) \equiv -\sqrt{2} \mathcal{D}_\mu \varphi^\alpha ~, \tag{3.4}$$

then its comparison with (3.3a) implies that

$$\xi^\alpha{}_I = -\tfrac{1}{\sqrt{2}} f_{IJ}{}^K V^{a\alpha} L_a{}^J L_K ~, \tag{3.5}$$

with the Killing vectors $\xi^\alpha{}_I$ in the directions of the gauged group $G$. Eq. (3.3c) has a new term for the non-Abelian coupling. Relevantly, we have

$$\mathcal{D}_\mu L_I = \mathcal{P}_\mu{}^a L_{Ia} ~, \quad \mathcal{D}_\mu L_I{}^a = \mathcal{P}_\mu{}^a L_I ~, \quad \mathcal{D}_\mu L^I = -\mathcal{P}_\mu{}^a L_a{}^I ~, \quad \mathcal{D}_\mu L_a{}^I = -\mathcal{P}_{\mu a} L^I ~. \tag{3.6}$$

By defining

$$C_{ab} \equiv f_{IJ}{}^K L_a{}^I L_b{}^J L_K = -\sqrt{2} \xi^\beta{}_I V_{\beta b} L_a{}^I = -C_{ba} ~, \tag{3.7}$$

we have the important relationship

$$\mathcal{D}_{[\mu}\mathcal{P}_{\nu]a} = +\tfrac{1}{2}\overline{g}F_{\mu\nu}{}^I L_I{}^b C_{ab} \quad , \tag{3.8}$$

by the use of another identity

$$L_I{}^b C_{ab} = -f_{IJ}{}^K L_a{}^J L_K \quad , \tag{3.9}$$

confirmed by (2.1). As has been already mentioned, in expressions like (3.7) - (3.9), the structure constants $f_{IJ}{}^K$ in any directions of the $U(1)$'s or $SO(2)$ are supposed to vanish, not to mention any other 'mixed' directions of different groups. The same is also true for the index $I$ in the last term in (3.3d), in which any irrelevant component gives the vanishing of $A_\mu{}^I$. These geometrical structures are parallel to the 7D case in [19].

Note that in this non-Abelian gauging, the gaugini $\lambda^{aA}$ are *not* in the adjoint representation, as opposed to the usual vector multiplets in higher dimensions [17], such as that in 10D with the gaugino in the adjoint representation. This is in a sense not surprising, because the gaugino fields should form the **n**-representation instead of the **(n + 1)**-representation of $SO(n,1)$, and therefore their range of indices should differ from that of the vector fields. This situation in 5D is similar to the original work in [6], or also in [9][19].

We next introduce the Killing vectors $\widehat{\widetilde{\xi}}^{\underline{\alpha}\hat{I}}$ for the direction of the gauged group $G$. In order to fix an invariant action, we also follow the result in [19], and we can postulate an additional term needed in the lagrangian [19]

$$e^{-1}\mathcal{L}_{\overline{g}} = +ia\overline{g}e^\sigma C_{ab}(\overline{\lambda}^a \lambda^b) \quad , \tag{3.10}$$

for the non-Abelian gauging, while putting *no explicit*[5] $\overline{g}$-dependent terms in the transformation rules. Now the variation of $\mathcal{L}_{\overline{g}}$ generates only two sorts of terms, when fermionic cubic terms are ignored: (i) $\overline{g}\lambda F$-terms and (ii) $\overline{g}\lambda\mathcal{D}\varphi$-terms. For the term in (i), the variation of the $\psi_\mu\lambda\mathcal{D}\varphi$ Noether-term is the only counter-contribution, while for the term in (ii), the kinetic term of $\varphi$ is the only contribution to cancel. Both of these two sectors yield the same condition $a = +2^{-3/2}$ consistently.

Armed with these preliminaries, we are ready to give the lagrangian

$$\begin{aligned} e^{-1}\mathcal{L}_1 = & -\tfrac{1}{4}R - \tfrac{1}{2}(\overline{\psi}_\mu \gamma^{\mu\nu\rho}D_\nu \psi_\rho) - \tfrac{1}{12}e^{-4\sigma}G_{\mu\nu\rho}^2 - \tfrac{1}{4}e^{-2\sigma}(L_I{}^a L_{Ja} + L_I L_J)F_{\mu\nu}{}^I F^{\mu\nu\,J} \\ & -\tfrac{1}{2}(\overline{\lambda}^a \gamma^\mu D_\mu \lambda_a) - \tfrac{1}{2}g_{\alpha\beta}(\mathcal{D}_\mu \varphi^\alpha)(\mathcal{D}^\mu \varphi^\beta) - \tfrac{3}{4}(\partial_\mu \sigma)^2 - \tfrac{1}{2}(\overline{\chi}\gamma^\mu D_\mu \chi) \\ & -\tfrac{1}{2}g_{\underline{\alpha}\underline{\beta}}(\partial_\mu \phi^{\underline{\alpha}})(\partial^\mu \phi^{\underline{\beta}}) - \tfrac{1}{2}(\overline{\psi}^{\underline{a}}\gamma^\mu D_\mu \psi_{\underline{a}}) \\ & +\tfrac{i}{\sqrt{2}}V_\alpha{}^a(\overline{\psi}_\mu \gamma^\nu \gamma^\mu \lambda_a)\mathcal{D}_\nu \varphi^\alpha + \tfrac{\sqrt{3}i}{2}(\overline{\psi}_\mu \gamma^\nu \gamma^\mu \chi)\partial_\nu \sigma + iV_{\underline{\alpha}}{}^{\underline{a}A}(\overline{\psi}_\mu \gamma^\nu \gamma^\mu \psi_{\underline{a}})\partial_\nu \phi^{\underline{\alpha}} \end{aligned}$$

---

[5]The word 'explicit' here implies any $\overline{g}$-dependent term other than those hidden in the covariant derivatives such as (3.3).

$$-\tfrac{i}{4\sqrt{2}}e^{-\sigma}(\overline{\psi}_\mu\gamma^{\mu\nu\rho\sigma}\psi_\nu + 2\overline{\psi}^\rho\psi^\sigma)L_I F_{\rho\sigma}{}^I + \tfrac{i}{6\sqrt{3}}e^{-2\sigma}(\overline{\psi}_\mu\gamma^{\rho\sigma\tau}\gamma^\mu\chi)G_{\rho\sigma\tau}$$

$$-\tfrac{1}{24}e^{-2\sigma}(\overline{\psi}_\mu\gamma^{\mu\nu\rho\sigma\tau}\psi_\nu - 6\overline{\psi}^\rho\gamma^\sigma\psi^\tau)G_{\rho\sigma\tau} - \tfrac{5}{72}e^{-2\sigma}(\overline{\chi}\gamma^{\mu\nu\rho}\chi)G_{\mu\nu\rho}$$

$$-\tfrac{1}{2\sqrt{2}}e^{-\sigma}(\overline{\psi}_\mu\gamma^{\rho\sigma}\gamma^\mu\lambda_a)L_I{}^a F_{\rho\sigma}{}^I - \tfrac{1}{2\sqrt{6}}e^{-\sigma}(\overline{\psi}_\mu\gamma^{\rho\sigma}\gamma^\mu\chi)L_I F_{\rho\sigma}{}^I - \tfrac{i}{12\sqrt{2}}e^{-\sigma}(\overline{\chi}\gamma^{\mu\nu}\chi)L_I F_{\mu\nu}{}^I$$

$$+\tfrac{i}{4\sqrt{2}}e^{-\sigma}(\overline{\lambda}^a\gamma^{\mu\nu}\lambda_a)L_I F_{\mu\nu}{}^I + \tfrac{1}{24}e^{-2\sigma}(\overline{\lambda}^a\gamma^{\mu\nu\rho}\lambda_a)G_{\mu\nu\rho} + \tfrac{1}{24}e^{-2\sigma}(\overline{\psi}^{\underline{a}}\gamma^{\mu\nu\rho}\psi_{\underline{a}})G_{\mu\nu\rho}$$

$$+\tfrac{i}{\sqrt{6}}e^{-2\sigma}(\overline{\chi}\gamma^{\mu\nu}\lambda_a)L_I{}^a F_{\mu\nu}{}^I - \tfrac{1}{4\sqrt{2}}e^{-\sigma}(\overline{\psi}^{\underline{a}}\gamma^{\mu\nu}\psi_{\underline{a}})L_I F_{\mu\nu}{}^I$$

$$+\tfrac{i}{2\sqrt{2}}\overline{g}e^\sigma C_{ab}(\overline{\lambda}^a\lambda^b)\ ,\tag{3.11}$$

with all of the $\mathcal{D}_\mu$ and $G_{\mu\nu\rho}$ in (3.3), yielding an invariant action $S_1$ under supersymmetry

$$\delta_Q e_\mu{}^m = +(\overline{\epsilon}\gamma^m\psi_\mu)\ ,\quad \delta_Q\sigma = +\tfrac{i}{\sqrt{3}}(\overline{\epsilon}\chi)\ ,$$

$$\delta_Q\psi_\mu{}^A = +D_\mu\epsilon^A + \tfrac{i}{6\sqrt{2}}e^{-\sigma}(\gamma_\mu{}^{\rho\sigma} - 4\delta_\mu{}^\rho\gamma^\sigma)\epsilon^A L_I F_{\rho\sigma}{}^I + \tfrac{1}{18}e^{-2\sigma}(\gamma_\mu{}^{\rho\sigma\tau} - \tfrac{3}{2}\delta_\mu{}^\rho\gamma^{\sigma\tau})\epsilon^A G_{\rho\sigma\tau}\ ,$$

$$\delta_Q A_\mu{}^I = -\tfrac{i}{\sqrt{2}}e^\sigma L^I(\overline{\epsilon}\psi_\mu) + \tfrac{1}{\sqrt{6}}e^\sigma L^I(\overline{\epsilon}\gamma_\mu\chi) + \tfrac{1}{\sqrt{2}}e^\sigma(\overline{\epsilon}\gamma_\mu\lambda^a)L_a{}^I\ ,$$

$$\delta_Q B_{\mu\nu} = +e^{2\sigma}(\overline{\epsilon}\gamma_{[\mu}\psi_{\nu]}) + \tfrac{i}{\sqrt{3}}e^{2\sigma}(\overline{\epsilon}\gamma_{\mu\nu}\chi) - 2L_{IJ}A_{[\mu|}{}^I(\delta_Q A_{|\nu]}{}^J)\ ,$$

$$\delta_Q\chi^A = -\tfrac{1}{2\sqrt{6}}e^{-\sigma}\gamma^{\mu\nu}\epsilon^A L_I F_{\mu\nu}{}^I + \tfrac{i}{6\sqrt{3}}e^{-2\sigma}\gamma^{\mu\nu\rho}\epsilon^A G_{\mu\nu\rho} - \tfrac{\sqrt{3}i}{2}\gamma^\mu\epsilon^A\partial_\mu\sigma\ ,$$

$$\delta_Q\varphi^\alpha = +\tfrac{i}{\sqrt{2}}V_a{}^\alpha(\overline{\epsilon}\lambda^a)\ ,$$

$$\delta_Q\lambda^{aA} = -\tfrac{1}{2\sqrt{2}}e^{-\sigma}\gamma^{\mu\nu}\epsilon^A L_I{}^a F_{\mu\nu}{}^I - \tfrac{i}{\sqrt{2}}\gamma^\mu\epsilon^A V_\alpha{}^a \mathcal{D}_\mu\varphi^\alpha\ ,$$

$$\delta_Q\phi^{\underline{\alpha}} = +iV_{\underline{a}A}{}^{\underline{\alpha}}(\overline{\epsilon}^A\psi^{\underline{a}})\ ,$$

$$\delta_Q\psi^{\underline{a}} = -iV_{\underline{\alpha}}{}^{\underline{a}A}\gamma^\mu\epsilon_A\partial_\mu\phi^{\underline{\alpha}}\ ,\tag{3.12}$$

Note that there is no need of any explicitly $\overline{g}$-dependent terms in the transformation rule. There is no potential term generated in this gauging, which is similar to the conventional $N = 2$ theories in 5D [7][9][8]. Compared with [19], since our vector fields do not carry extra $Sp(1)$ indices, no scalar potential term is generated.

Analogous to (2.5), we have the scaling invariance of $\mathcal{L}_1$ when the coupling constant $\overline{g}$ transforms as

$$\overline{g} \to e^{-c}\overline{g}\ ,\tag{3.13}$$

when the fields transform as in (2.5).

## 4. $Sp(n') \times Sp(1)$-Gauging

In our previous paper [12], we studied the gauging of $SO(2)$ which is the subgroup of $Sp(1)$ in the isotropy groups $Sp(n') \times Sp(1)$ in the coset $Sp(n',1)/Sp(n') \times Sp(1)$. Most of the geometric relationships related to the coset $SO(n,1)/SO(n)$ are parallel to the $SO(2)$-gauging [12], so we give important relations in such a way that the comparison with [12] is easy to make.

Our total gauged group in this section is $G = SO(2) \times Sp(n') \times Sp(1) \times \overline{H} \times [U(1)]^{n-p+1}$, which is a special case of the previous section. In fact, the first $SO(2)$ is for the $I = 0$-direction for the indices $I = 0, 1, \cdots, n+1$, and the groups $Sp(n') \times Sp(1)$ are regarded as a special case of $H \equiv Sp(n') \times Sp(1) \times \overline{H}$ for the group $H$ in the last section, and an arbitrary gauge group $\overline{H}$ with $\dim \overline{H} = p - n'(2n'+1) - 4$, such that the previous condition $\dim H = p - 1$ is maintained. Since the dimension $p$ is still arbitrary, we have enough freedom for choosing the group $\overline{H}$ for a large enough dimension of $n$.

Accordingly, we arrange our index convention as follows. Among the indices $I, J, \cdots = 0, 1, 2, \cdots, n$ for the total $n+1$ copies of vector fields, we use the indices $\underline{I}, \underline{J}, \cdots = 1, 2, \cdots, n'(2n'+1)$ for the adjoint representation of $Sp(n')$, and combine them with $i, j, \cdots = 1, 2, 3$ for that of $Sp(1)$, in terms of the combined indices $\hat{\imath} \equiv (\underline{I}, i), \hat{\jmath} \equiv (\underline{J}, j), \cdots$ for the gauged groups $Sp(n') \times Sp(1)$. For the adjoint indices for $\overline{H}$, we use the *barred* ones: $\overline{I}, \overline{J}, \cdots = 1, 2, \cdots, p-1$. As for the remaining product groups $SO(2) \times [U(1)]^{n-p+1}$, we do not need particular indices in this section, so we do not specify the indices for these groups. Compared with the indices $I, J, \cdots$, all these indices $\hat{\imath}, \hat{\jmath}, \cdots; \underline{I}, \underline{J}, \cdots; i, j, \cdots; \overline{I}, \overline{J}, \cdots$ do not need distinctions of their raising/lowering due to their positive definite metrics. Therefore their contractions are given as superscripts like $A^{\underline{I}} B^{\underline{I}}$.

In our previous paper [6], the $SO(2)$-gauging was performed by introducing the constant vectors $V^I$, with the coupling constant $g$. In our present case of $Sp(n') \times Sp(1)$-gauging, this $SO(2)$ group is enlarged to $Sp(n') \times Sp(1)$. In this section, we use the coupling constant $g$ for $Sp(1)$, $g'$ for $Sp(n')$, and $\overline{g}$ for $\overline{H}$. Accordingly, all the combination of $gV_I A_\mu{}^I \xi^{\underline{\alpha}}$ in [12] will be replaced by $gA_\mu{}^i \xi^{\underline{\alpha}i} + g'A_\mu{}^{\underline{I}} \xi^{\underline{\alpha}\underline{I}}$, where $\xi^{\underline{\alpha}\underline{I}}$ and $\xi^{\underline{\alpha}i}$ are the Killing vectors for the gauged groups $Sp(n') \times Sp(1)$ in the coset $Sp(n', 1)/Sp(n') \times Sp(1)$.

Accordingly, the covariant derivatives on $Sp(1)$ non-invariant fermions acquire the $Sp(1)$ minimal couplings in addition to the $D_\mu$'s or $\partial_\mu \varphi^\alpha$ in section 2 as

$$D_\mu \epsilon^A \longrightarrow \mathcal{D}_\mu \epsilon^A \equiv D_\mu(\omega)\epsilon^A + (\mathcal{D}_\mu \phi^{\underline{\alpha}}) A_{\underline{\alpha}}{}^i (T^i \epsilon)^A + gA_\mu{}^i (T^i \epsilon)^A , \tag{4.1a}$$

$$D_{[\mu} \psi_{\nu]}{}^A \longrightarrow \mathcal{D}_{[\mu} \psi_{\nu]}{}^A \equiv D_{[\mu}(\omega)\psi_{\nu]}{}^A + (\mathcal{D}_{[\mu} \phi^{\underline{\alpha}}) A_{\underline{\alpha}}{}^i (T^i \psi_{\nu]})^A + gA_{[\mu}{}^i (T^i \psi_{\nu]})^A , \tag{4.1b}$$

$$D_\mu \chi^A \longrightarrow \mathcal{D}_\mu \chi^A \equiv D_\mu(\omega)\chi^A + (\mathcal{D}_\mu \phi^{\underline{\alpha}}) A_{\underline{\alpha}}{}^i (T^i \chi)^A + gA_\mu{}^i (T^i \chi)^A , \tag{4.1c}$$

$$D_\mu \lambda^{aA} \longrightarrow \mathcal{D}_\mu \lambda^{aA} \equiv D_\mu(\omega)\lambda^{aA} + \mathcal{A}_\mu{}^{ab} \lambda_b{}^A + (\mathcal{D}_\mu \phi^{\underline{\alpha}}) A_{\underline{\alpha}}{}^i (T^i \lambda^a)^A + gA_\mu{}^i (T^i \lambda^a)^A \tag{4.1d}$$

$$\partial_\mu \varphi^\alpha \longrightarrow \mathcal{D}_\mu \varphi^\alpha \equiv \partial_\mu \varphi^\alpha - A_\mu{}^I \widehat{\xi}^\alpha{}_I , \tag{4.1e}$$

$$D_\mu \psi^{\underline{a}} \longrightarrow \mathcal{D}_\mu \psi^{\underline{a}} \equiv D_\mu(\omega)\psi^{\underline{a}} + (\mathcal{D}_\mu \phi^{\underline{\alpha}}) A_{\underline{\alpha}}{}^{\underline{I}} (T^{\underline{I}} \psi)^{\underline{a}} + g'A_\mu{}^{\underline{I}} (T^{\underline{I}} \psi)^{\underline{a}} , \tag{4.1f}$$

$$\partial_\mu \phi^{\underline{\alpha}} \longrightarrow \mathcal{D}_\mu \phi^{\underline{\alpha}} \equiv \partial_\mu \phi^{\underline{\alpha}} - g'A_\mu{}^{\underline{I}} \xi^{\underline{\alpha}\underline{I}} - gA_\mu{}^i \xi^{\underline{\alpha}i} \equiv \partial_\mu \phi^{\underline{\alpha}} - A_\mu{}^{\hat{\imath}} \widehat{\xi}^{\underline{\alpha}\hat{\imath}} , \tag{4.1g}$$

with the generalized Killing vectors

$$\widehat{\xi}^{\underline{\alpha}}{}_I \equiv \begin{cases} \widehat{\xi}^{\underline{\alpha}\hat{I}} \equiv \begin{cases} \widehat{\xi}^{\underline{\alpha}I} \equiv g'\xi^{\underline{\alpha}I} & (\text{for } Sp(n')) \ , \\ \widehat{\xi}^{\underline{\alpha}i} \equiv g\xi^{\underline{\alpha}i} & (\text{for } Sp(1)) \ , \end{cases} \\ 0 & (\text{otherwise}) \ . \end{cases} \quad (4.2)$$

For example, an expression like $\widehat{\xi}^{\underline{\alpha}}{}_I L^I$ actually means $\widehat{\xi}^{\underline{\alpha}}{}_I L^I \equiv \widehat{\xi}^{\underline{\alpha}\hat{I}} L^{\hat{I}} \equiv g'\xi^{\underline{\alpha}I} L^{\underline{I}} + g\xi^{\underline{\alpha}i} L^i$. The absence of the component $\widehat{\xi}^{\underline{\alpha}\overline{J}}$ is understood from the fact that the group $\overline{H}$ has nothing to do with the coset $Sp(n',1)/Sp(n') \times Sp(1)$. In (4.1), all the terms other than explicit $g$-terms are just the previous covariant derivatives in (2.6) in which $\partial_\mu \phi^{\underline{\alpha}}$ is replaced by $\mathcal{D}_\mu \phi^{\underline{\alpha}}$, and the matrices $T^{\hat{I}}$ are the anti-hermitian generator of $Sp(n') \times Sp(1)$, as its index $\hat{i}$ reveals. This structure is similar to the models in [6][7][9][8].

The covariance of the derivatives in (4.1) are confirmed by considering the transformations of these fields under the gauged groups $G \equiv SO(2) \times Sp(n') \times Sp(1) \times \overline{H} \times [\,U(1)\,]^{n-p+1}$, such as

$$\delta_G \phi^{\underline{\alpha}} = +\alpha^I \widehat{\xi}^{\underline{\alpha}}{}_I \ , \quad \delta_G A_\mu{}^I = \partial_\mu \alpha^I + \widehat{f}_{JK}{}^I A_\mu{}^J \alpha^K \equiv D_\mu \alpha^I \ , \quad (4.3a)$$

$$\delta_G \widehat{\xi}^{\underline{\alpha}}{}_I = \alpha^J \widehat{\xi}^{\underline{\beta}}{}_I (\partial_{\underline{\beta}} \widehat{\xi}^{\underline{\alpha}}{}_J) - \widehat{f}_{IJ}{}^K \alpha^J \widehat{\xi}^{\underline{\alpha}}{}_K \ , \quad (4.3b)$$

$$\delta_G (\mathcal{D}_\mu \phi^{\underline{\alpha}}) = \alpha^I (\mathcal{D}_\mu \phi^{\underline{\alpha}})(\partial_{\underline{\beta}} \widehat{\xi}^{\underline{\alpha}}{}_I) \ , \quad (4.3c)$$

for the local parameters $\alpha^I$ for the gauged groups in $G$, and the structure constants

$$\widehat{f}_{IJ}{}^K \equiv \begin{cases} \widehat{f}^{\hat{I}\hat{J}\hat{K}} = \begin{cases} \widehat{f}^{\underline{IJK}} = g' f^{\underline{IJK}} & (\text{for } Sp(n')) \ , \\ \widehat{f}^{ijk} = g\epsilon^{ijk} & (\text{for } Sp(1)) \ , \end{cases} \\ \widehat{f}^{\overline{IJK}} = \overline{g} f^{\overline{IJK}} & (\text{for } \overline{H}) \ , \\ 0 & (\text{otherwise}) \ , \end{cases} \quad (4.4a)$$

$$\widehat{f}_{IJK} \equiv \widehat{f}_{IJ}{}^L L_{LK} = \widehat{f}_{[IJK]} \ , \quad (4.4b)$$

for the respective structure constants for $Sp(n')$, $Sp(1)$ and $\overline{H}$ in the combined notation. Since the $SO(2)$ group in the negative metric 0-th direction is Abelian, it does not enter (4.4), and therefore we do not need to distinguish the super/subscripts on the r.h.s. of (4.4a).

The field strength (3.3d) should be also modified by all the non-Abelian couplings:

$$G_{\mu\nu\rho} \equiv 3\left(\partial_{[\mu} B_{\nu\rho]} - L_{IJ} F_{[\mu\nu}{}^I A_{\rho]}{}^J + \tfrac{1}{3} \widehat{f}_{IJK} A_\mu{}^I A_\nu{}^J A_\rho{}^K\right) \ . \quad (4.5)$$

The commutator of two covariant derivatives acting on $\epsilon^A$ provides certain important geometric quantity in our system:

$$[\mathcal{D}_\mu, \mathcal{D}_\nu]\epsilon_A = -\tfrac{1}{4} R_{\mu\nu}{}^{mn} \gamma_{mn} \epsilon_A + (\mathcal{D}_{[\mu|} \phi^{\underline{\alpha}})(\mathcal{D}_{|\nu]} \phi^{\underline{\beta}}) F_{\underline{\alpha\beta}}{}^i (T^i \epsilon)_A - F_{\mu\nu}{}^{\hat{I}} \widehat{C}^{i\hat{I}} (T^i \epsilon)_A \ , \quad (4.6)$$

where the function $\widehat{C}^{i\hat{I}}$ is defined by

$$\widehat{C}^{iJ} \equiv \begin{cases} \widehat{C}^{i\hat{J}} = \begin{cases} \widehat{C}^{i\underline{J}} \equiv g'C^{i\underline{J}} \equiv g'A_{\underline{\alpha}}{}^i \xi^{\underline{\alpha}J} & \text{(for} \quad Sp(n')) \\ \widehat{C}^{ij} \equiv gC^{ij} \equiv g(A_{\underline{\alpha}}{}^i \xi^{\underline{\alpha}j} - \delta^{ij}) & \text{(for} \quad Sp(1)) \end{cases} \\ 0 \quad \text{(otherwise)} \end{cases}, \tag{4.7}$$

which is analogous to the $N = 2$ case in 6D [16], or our combination $\widehat{C}^{i\hat{J}}T^i$ is an analog of $P_{Ii}{}^j$ in the notation in [9]. The component $\widehat{C}^{ij}$ in (4.7) implies that all the terms with $gT^2V_I$ in [12] should be replaced by $-T^i\widehat{C}^i{}_I$, when we gauge $G \equiv SO(2) \times Sp(n') \times Sp(1) \times \overline{H} \times [U(1)]^{n-p+1}$ instead of $SO(2)$ in [12]. Some illustrative examples of the replacements of the terms in [12] are given by

$$+gT^2V_I \longrightarrow -T^i\widehat{C}^i{}_I \ , \tag{4.8a}$$

$$+g\xi^{\underline{\alpha}}V_I \longrightarrow +\widehat{\xi}^{\underline{\alpha}}{}_I \ , \tag{4.8b}$$

$$+\tfrac{i}{2\sqrt{2}}ge^\sigma(\overline{\psi}_\mu \gamma^{\mu\nu} T_2 \psi_\nu) V_I L^I \longrightarrow -\tfrac{i}{2\sqrt{2}}e^\sigma(\overline{\psi}_\mu \gamma^{\mu\nu} T_i \psi_\nu) \widehat{C}^i{}_I L^I \ , \tag{4.8c}$$

$$-\tfrac{1}{8}g^2 e^{2\sigma} V_I V_J L^{IJ} \longrightarrow -\tfrac{1}{8}\widehat{C}^i{}_I \widehat{C}^i{}_J L^{IJ} \ . \tag{4.8d}$$

Needless to say, when the gauged group is truncated from $G \equiv SO(2) \times Sp(n') \times Sp(1) \times \overline{H} \times [U(1)]^{n-p+1}$ back into $SO(2)$ with the $V_I$'s as in [12], then all the r.h.s. in (4.8) go back to their l.h.s. This can provide a good confirmation at various stages of computations, in particular the invariance check of total action under supersymmetry. Due to the indefinite metric involved, special care is needed for the contraction of the $I$-indices here, while the ups/downs of the index $i$ does not matter. Relevantly, we can define the covariant derivative on $\widehat{C}^{i\hat{I}}$ as

$$D_{\underline{\alpha}} \widehat{C}^{i\hat{J}} \equiv \partial_{\underline{\alpha}} \widehat{C}^{i\hat{J}} + \epsilon^{ijk} A_{\underline{\alpha}}{}^j \widehat{C}^{k\hat{J}} \ , \tag{4.9}$$

so that

$$\mathcal{D}_\mu \widehat{C}^{i\hat{I}} \equiv \partial_\mu \widehat{C}^{i\hat{I}} + g\epsilon^{ijk} A_\mu{}^j \widehat{C}^{k\hat{I}} + \widehat{f}^{\hat{I}\hat{J}\hat{K}} A_\mu{}^{\hat{J}} \widehat{C}^{i\hat{K}} + \epsilon^{ijk} (\mathcal{D}_\mu \phi^{\underline{\alpha}}) A_{\underline{\alpha}}{}^j \widehat{C}^{k\hat{I}}$$

$$= (\mathcal{D}_\mu \phi^{\underline{\alpha}})(D_{\underline{\alpha}} \widehat{C}^{i\hat{I}}) \ . \tag{4.10}$$

To confirm the last equality, we need the relationship

$$\widehat{\xi}^{\underline{\alpha}\hat{I}} \partial_{\underline{\alpha}} \widehat{C}^{i\hat{J}} = \widehat{f}^{\hat{I}\hat{J}\hat{K}} \widehat{C}^{i\hat{K}} + \widehat{f}^{\hat{I}ik} \widehat{C}^{k\hat{J}} \ , \tag{4.11}$$

derived from the Lie derivatives

$$\mathcal{L}_{\widehat{\xi}^{\hat{I}}} \widehat{\xi}^{\underline{\alpha}\hat{J}} \equiv \widehat{\xi}^{\underline{\beta}\hat{I}} \partial_{\underline{\beta}} \widehat{\xi}^{\underline{\alpha}\hat{J}} - \widehat{\xi}^{\underline{\beta}\hat{J}} \partial_{\underline{\beta}} \widehat{\xi}^{\underline{\alpha}\hat{I}} = \widehat{f}^{\hat{I}\hat{J}\hat{K}} \widehat{\xi}^{\underline{\alpha}\hat{K}} \ ,$$

$$\mathcal{L}_{\widehat{\xi}^{\hat{I}}} A_{\underline{\alpha}}{}^i \equiv \widehat{\xi}^{\underline{\beta}\hat{I}} \partial_{\underline{\beta}} A_{\underline{\alpha}}{}^i - (\partial_{\underline{\alpha}} \widehat{\xi}^{\underline{\beta}\hat{I}}) A_{\underline{\beta}}{}^i = \widehat{f}^{\hat{I}ij} A_{\underline{\alpha}}{}^j \ . \tag{4.12}$$

Note that as in the 6D case in [16], there is no term with $\widehat{f}^{\hat{J}\hat{K}\hat{L}}A_{\underline{\alpha}}{}^{\hat{K}}$ needed in (4.9), in order to be consistent with supersymmetry of the action. Other important corollaries with these $\widehat{C}$'s are such as

$$D_{\underline{\alpha}}\widehat{C}^{i\hat{I}} = \widehat{\xi}^{\underline{\beta}\hat{I}}F_{\underline{\alpha\beta}}{}^i \ , \quad \widehat{\xi}^{\underline{\alpha}\hat{I}}D_{\underline{\alpha}}\widehat{C}^{i\hat{I}} \equiv 0 \ , \quad (4.13)$$

which have parallel structures as in the 6D case [16].

The most crucial relationship involving $\widehat{C}^{iJ}$ in our system is the constraint required by the supersymmetric invariance of the total action, needed for the consistency between the two cosets $SO(n,1)/SO(n)$ and $Sp(n',1)/Sp(n') \times Sp(1)$:

$$F_{\underline{\alpha\beta}}{}^i\widehat{\xi}^{\underline{\alpha}}{}_I\widehat{\xi}^{\underline{\beta}}{}_J - \epsilon^{ijk}\widehat{C}^j{}_I\widehat{C}^k{}_J - \widehat{f}_{IJ}{}^K\widehat{C}^i{}_K = 0 \ . \quad (4.14)$$

This constraint is required by the cancellation of $\lambda$-linear terms with the structure $(\overline{\epsilon}T^i\lambda)$ with one $T^i$-generator sandwiched. This constraint is also analogous to eq. (3.15) in [9], or to eqs. (2.21) - (2.24) in [8]. The necessity of such a constraint is natural from the fact that the vector fields $A_\mu{}^I$ in our system are *both* in the $(\mathbf{n+1})$-representation of $SO(n,1)$ *and* the adjoint representations of the gauged groups in $G$ at the same time. And therefore their mutual consistency, in particular, under supersymmetry requires such a constraint. It is taken for granted that in (4.14), there are many trivially vanishing components for each terms depending on the combination of the adjoint indices. For example, according to (4.4), the structure constants $\widehat{f}_{IJ}{}^K$ vanishes identically for any $U(1)$-directions, or for any 'mixed' directions of different gauge groups. However, note that the first term in (4.14) does not automatically vanish for such 'mixed' directions. Our previous $SO(2)$ gauging in [12] also satisfies (4.14) trivially, because the last two terms vanish, while $\widehat{\xi}^{\underline{\alpha}}{}_I \to g\xi^{\underline{\alpha}}V_I$ makes the first term vanish, too.

The tensor $C_{ab}$ in (3.7) is also redefined in terms of $\widehat{f}_{IJ}{}^K$ by

$$\widehat{C}_{ab} \equiv \widehat{f}_{IJ}{}^K L_a{}^I L_b{}^J L_K = -\widehat{C}_{ba} \ . \quad (4.15)$$

With these preliminaries, we now give our lagrangian[6]

$$e^{-1}\mathcal{L}_2 = -\tfrac{1}{4}R - \tfrac{1}{2}(\overline{\psi}_\mu\gamma^{\mu\nu\rho}\mathcal{D}_\nu\psi_\rho) - \tfrac{1}{12}e^{-4\sigma}G^2_{\mu\nu\rho} - \tfrac{1}{4}e^{-2\sigma}(L_I{}^a L_{Ja} + L_I L_J)F_{\mu\nu}{}^I F^{\mu\nu J}$$
$$-\tfrac{1}{2}(\overline{\lambda}^a\gamma^\mu\mathcal{D}_\mu\lambda_a) - \tfrac{1}{2}g_{\alpha\beta}(\mathcal{D}_\mu\varphi^\alpha)(\mathcal{D}^\mu\varphi^\beta) - \tfrac{3}{4}(\partial_\mu\sigma)^2 - \tfrac{1}{2}(\overline{\chi}\gamma^\mu\mathcal{D}_\mu\chi)$$
$$-\tfrac{1}{2}g_{\underline{\alpha\beta}}(\mathcal{D}_\mu\phi^{\underline{\alpha}})(\mathcal{D}^\mu\phi^{\underline{\beta}}) - \tfrac{1}{2}(\overline{\psi}^{\underline{a}}\gamma^\mu\mathcal{D}_\mu\psi_{\underline{a}})$$
$$+\tfrac{i}{\sqrt{2}}V_\alpha{}^a(\overline{\psi}_\mu\gamma^\nu\gamma^\mu\lambda_a)\mathcal{D}_\nu\varphi^\alpha + \tfrac{\sqrt{3}i}{2}(\overline{\psi}_\mu\gamma^\nu\gamma^\mu\chi)\partial_\nu\sigma + iV_{\underline{\alpha}}{}^{\underline{a}A}(\overline{\psi}_\mu\gamma^\nu\gamma^\mu\psi_{\underline{a}})\mathcal{D}_\nu\phi^{\underline{\alpha}}$$

---

[6]We mention the errors in signatures of terms in our previous paper [12]. The sign errors in the $g$-linear lagrangian (4.2) and $g$-linear transformation rule (4.3) in [12] are now corrected in (4.16) and (4.17).

$$\begin{aligned}
&- \tfrac{i}{4\sqrt{2}} e^{-\sigma} (\overline{\psi}_\mu \gamma^{\mu\nu\rho\sigma} \psi_\nu + 2\overline{\psi}^\rho \psi^\sigma) L_I F_{\rho\sigma}{}^I + \tfrac{i}{6\sqrt{3}} e^{-2\sigma} (\overline{\psi}_\mu \gamma^{\rho\sigma\tau} \gamma^\mu \chi) G_{\rho\sigma\tau} \\
&- \tfrac{1}{24} e^{-2\sigma} (\overline{\psi}_\mu \gamma^{\mu\nu\rho\sigma\tau} \psi_\nu - 6\overline{\psi}^\rho \gamma^\sigma \psi^\tau) G_{\rho\sigma\tau} - \tfrac{5}{72} e^{-2\sigma} (\overline{\chi} \gamma^{\mu\nu\rho} \chi) G_{\mu\nu\rho} \\
&- \tfrac{1}{2\sqrt{2}} e^{-\sigma} (\overline{\psi}_\mu \gamma^{\rho\sigma} \gamma^\mu \lambda_a) L_I{}^a F_{\rho\sigma}{}^I - \tfrac{1}{2\sqrt{6}} e^{-\sigma} (\overline{\psi}_\mu \gamma^{\rho\sigma} \gamma^\mu \chi) L_I F_{\rho\sigma}{}^I - \tfrac{i}{12\sqrt{2}} e^{-\sigma} (\overline{\chi} \gamma^{\mu\nu} \chi) L_I F_{\mu\nu}{}^I \\
&+ \tfrac{i}{4\sqrt{2}} e^{-\sigma} (\overline{\lambda}^a \gamma^{\mu\nu} \lambda_a) L_I F_{\mu\nu}{}^I + \tfrac{1}{24} e^{-2\sigma} (\overline{\lambda}^a \gamma^{\mu\nu\rho} \lambda_a) G_{\mu\nu\rho} + \tfrac{1}{24} e^{-2\sigma} (\overline{\psi}^{\underline{a}} \gamma^{\mu\nu\rho} \psi_{\underline{a}}) G_{\mu\nu\rho} \\
&+ \tfrac{i}{\sqrt{6}} e^{-2\sigma} (\overline{\chi} \gamma^{\mu\nu} \lambda_a) L_I{}^a F_{\mu\nu}{}^I - \tfrac{1}{4\sqrt{2}} e^{-\sigma} (\overline{\psi}^{\underline{a}} \gamma^{\mu\nu} \psi_{\underline{a}}) L_I F_{\mu\nu}{}^I \\
&+ \Big[ - \tfrac{i}{2\sqrt{2}} e^{\sigma} (\overline{\psi}_\mu \gamma^{\mu\nu} T^i \psi_\nu) \widehat{C}^i{}_I L^I + \tfrac{1}{\sqrt{2}} e^{\sigma} (\overline{\psi}_\mu \gamma^\mu T^i \lambda^a) \widehat{C}^i{}_I L_a{}^I \\
&\qquad + \tfrac{i}{2\sqrt{2}} e^{\sigma} (\overline{\lambda}^a T^i \lambda_a) \widehat{C}^i{}_I L^I + \tfrac{2i}{\sqrt{6}} e^{\sigma} (\overline{\chi} T^i \lambda^a) \widehat{C}^i{}_I L_a{}^I \\
&\qquad - \tfrac{i}{6\sqrt{2}} e^{\sigma} (\overline{\chi} T^i \chi) \widehat{C}^i{}_I L^I + \tfrac{1}{\sqrt{6}} e^{\sigma} (\overline{\psi}_\mu \gamma^\mu T^i \chi) \widehat{C}^i{}_I L^I \\
&\qquad + \sqrt{2} i e^{\sigma} (\overline{\psi}^{\underline{a}} \lambda^{aA}) V_{\underline{\alpha} a A} \widehat{\xi}^{\underline{\alpha}}{}_I L_a{}^I + \tfrac{1}{\sqrt{2}} e^{\sigma} (\overline{\psi}_\mu{}^A \gamma^\mu \psi^{\underline{a}}) V_{\underline{\alpha} a A} \widehat{\xi}^{\underline{\alpha}}{}_I L^I \\
&\qquad + \tfrac{2i}{\sqrt{6}} e^{\sigma} (\overline{\chi}^A \psi^{\underline{a}}) V_{\underline{\alpha} a A} \widehat{\xi}^{\underline{\alpha}}{}_I L^I + \tfrac{i}{4\sqrt{2}} e^{\sigma} (\overline{\psi}^{\underline{a}} \psi^{\underline{b}}) V_{\underline{a}}{}^{B\underline{\alpha}} V_{\underline{b} B} (D_{\underline{\alpha}} \widehat{\xi}^{\underline{\beta}}{}_I) L^I + \tfrac{i}{2\sqrt{2}} e^{\sigma} \widehat{C}_{ab} (\overline{\lambda}^a \lambda^b) \Big] \\
&+ \Big[ - \tfrac{1}{8} e^{2\sigma} \widehat{C}^{iI} \widehat{C}^{iJ} L_{IJ} - \tfrac{1}{4} e^{2\sigma} \widehat{\xi}_{\underline{\alpha}}{}^I \widehat{\xi}^{\underline{\alpha} J} L_I L_J \Big] \ , 
\end{aligned} \qquad (4.16)$$

where the penultimate pair of the square brackets [ ] is for the terms at $\mathcal{O}(\widetilde{g})$, while the last pair is for the terms at $\mathcal{O}(\widetilde{g}^2)$, where $\widetilde{g}$ is any minimal coupling constant for gaugings among $g$, $g'$ or $\overline{g}$. This is because $\widehat{C}^{iI}$, $\widehat{C}_{ab}$ and $\widehat{\xi}^{\underline{\alpha} I}$ are all at $\mathcal{O}(\widetilde{g})$. Our lagrangian (4.16) yields an action $S_2$ invariant under supersymmetry

$$\begin{aligned}
\delta_Q e_\mu{}^m &= +(\overline{\epsilon} \gamma^m \psi_\mu) \ , \qquad \delta_Q \sigma = +\tfrac{i}{\sqrt{3}} (\overline{\epsilon} \chi) \ , \\
\delta_Q \psi_\mu{}^A &= +\mathcal{D}_\mu \epsilon^A + \tfrac{i}{6\sqrt{2}} e^{-\sigma} (\gamma_\mu{}^{\rho\sigma} - 4\delta_\mu{}^\rho \gamma^\sigma) \epsilon^A L_I F_{\rho\sigma}{}^I + \tfrac{1}{18} e^{-2\sigma} (\gamma_\mu{}^{\rho\sigma\tau} - \tfrac{3}{2} \delta_\mu{}^\rho \gamma^{\sigma\tau}) \epsilon^A G_{\rho\sigma\tau} \\
&\quad - \tfrac{i}{3\sqrt{2}} e^\sigma (\gamma_\mu T^i \epsilon)^A \widehat{C}^i{}_I L^I \ , \\
\delta_Q A_\mu{}^I &= - \tfrac{i}{\sqrt{2}} e^\sigma L^I (\overline{\epsilon} \psi_\mu) + \tfrac{1}{\sqrt{6}} e^\sigma L^I (\overline{\epsilon} \gamma_\mu \chi) + \tfrac{1}{\sqrt{2}} e^\sigma (\overline{\epsilon} \gamma_\mu \lambda^a) L_a{}^I \ , \\
\delta_Q B_{\mu\nu} &= + e^{2\sigma} (\overline{\epsilon} \gamma_{[\mu} \psi_{\nu]}) + \tfrac{i}{\sqrt{3}} e^{2\sigma} (\overline{\epsilon} \gamma_{\mu\nu} \chi) - 2 L_{IJ} A_{[\mu}{}^I (\delta_Q A_{|\nu]}{}^J) \ , \\
\delta_Q \chi^A &= -\tfrac{1}{2\sqrt{6}} e^{-\sigma} \gamma^{\mu\nu} \epsilon^A L_I F_{\mu\nu}{}^I + \tfrac{i}{6\sqrt{3}} e^{-2\sigma} \gamma^{\mu\nu\rho} \epsilon^A G_{\mu\nu\rho} - \tfrac{\sqrt{3}i}{2} \gamma^\mu \epsilon^A \partial_\mu \sigma + \tfrac{1}{\sqrt{6}} e^\sigma (T^i \epsilon)^A \widehat{C}^i{}_I L^I \ , \\
\delta_Q \varphi^\alpha &= +\tfrac{i}{\sqrt{2}} V_a{}^\alpha (\overline{\epsilon} \lambda^a) \ , \\
\delta_Q \lambda^{aA} &= -\tfrac{1}{2\sqrt{2}} e^{-\sigma} \gamma^{\mu\nu} \epsilon^A L_I{}^a F_{\mu\nu}{}^I - \tfrac{i}{\sqrt{2}} \gamma^\mu \epsilon^A V_\alpha{}^a \partial_\mu \varphi^\alpha + \tfrac{1}{\sqrt{2}} e^\sigma (T^i \epsilon)^A \widehat{C}^i{}_I L^{aI} \ , \\
\delta_Q \phi^{\underline{\alpha}} &= +i V_{\underline{a} A}{}^{\underline{\alpha}} (\overline{\epsilon}^A \psi^{\underline{a}}) \ , \\
\delta_Q \psi^{\underline{a}} &= -i V_{\underline{\alpha}}{}^{\underline{a} A} \gamma^\mu \epsilon_A \mathcal{D}_\mu \phi^{\underline{\alpha}} - \tfrac{1}{\sqrt{2}} e^\sigma \epsilon_B V_{\underline{\beta}}{}^{\underline{a} B} \widehat{\xi}^{\underline{\beta}}{}_I L^I \ ,
\end{aligned} \qquad (4.17)$$

Similarly to the $SO(2)$-gauging [12], the potential term is positive definite, *except for* the term with $L_I \equiv L_I{}^{(0)}$ in $L_{IJ} = L_{aI} L_J{}^a - L_I L_J$ [6][7][9][8][12]:

$$V_{\mathrm{pot}} = +\tfrac{1}{8} e^{2\sigma} \widehat{C}^{iI} \widehat{C}^{iJ} L_{IJ} + \tfrac{1}{4} e^{2\sigma} \widehat{\xi}_{\underline{\alpha}}{}^I \widehat{\xi}^{\underline{\alpha} J} L_I L_J \ . \qquad (4.18)$$

As in section 3, our lagrangian $\mathcal{L}_2$ has the scaling invariance when $g$, $g'$ and $\overline{g}$ transform like

$$g \to e^{-c}g \ , \quad g' \to e^{-c}g' \ , \quad \overline{g} \to e^{-c}\overline{g} \ ,$$

$$\widehat{C}^{iI} \to e^{-c}\widehat{C}^{iI} \ , \quad \widehat{\xi}^{\underline{\alpha}I} \to e^{-c}\widehat{\xi}^{\underline{\alpha}I} \ , \quad \widehat{C}_{ab} \to e^{-c}\widehat{C}_{ab} \ , \tag{4.19}$$

in addition to (2.5).

Similarly to the 6D case with the $Sp(n') \times Sp(1)$-gauging [16], any subgroup of these gauge groups can be also gauged consistently with supersymmetry, even though the details of its process are skipped here. In such a case, the indices $I, J, \cdots$ and $i, j, k$ are to be replaced by the corresponding indices of such gauged subgroups. In particular, for the $SO(2)$ subgroup gauging out of the $Sp(1)$ above, only the second direction $2$ out of the original indices $i, j, k$ is relevant, so that we can use the notation such as $T_2$ as in [12]. As the $SO(2)$-gauging described in [12] or in [6][7][9][8], we can combine the products of $U(1)$ groups, by introducing the constant couplings $V_I$ $(I, J, \cdots = 0, 1, 2, \cdots, n)$, as a slight generalization of the single $SO(2)$ subgroup gauging. In any of these cases, our results above are formally valid, and only the interpretation or the range of indices are changed.

## 5. Alternative $N = 2$ Supergravity in Singular 5D Space-Time

As has been developed in [1][2] for the conventional $N = 2$ supergravity [6][7][9], we can generalize our alternative $N = 2$ supergravity into singular 5D space-time, as supersymmetrization of Randall-Sundrum brane-world scenario [3].

As in our previous paper [12], we follow the prescription in [2] designed for the case of Abelian $SO(2)$ gauging for the singular 5D space-time with the orbifold-type singularity of $S^1/\mathbb{Z}_2$. However, since our present total gauged group is non-Abelian: $G = SO(2) \times Sp(n') \times Sp(1) \times \overline{H} \times [U(1)]^{n-p+1}$ which is much bigger than just $SO(2)$, we need special care when applying the method in [2].

We start with fixing the bulk 5D space-time action $S_{\text{bulk}}$ before considering the singularity. Mimicking the Abelian case [2], we first replace the original $Sp(1)$-gauging coupling constant $g$ *everywhere* in $\mathcal{L}_2$ by a space-time-dependent real scalar field $G(x)$, and then introduce a fourth-rank antisymmetric tensor potential $A_{\mu\nu\rho\sigma}$, with a new term in the lagrangian [2]

$$S_{AG} \equiv \int d^5x \, \mathcal{L}_{AG} \equiv \int d^5x \left( \tfrac{1}{24} \epsilon^{\mu\nu\rho\sigma\tau} A_{\mu\nu\rho\sigma} \partial_\tau G \right) \ . \tag{5.1}$$

The reason we replace only $g$ by $G(x)$ is that this coupling is for the $Sp(1)$ group that can contain the $SO(2)$ subgroup in our previous case [12] which is analogous to the Abelian group in [2]. The scalar field $G(x)$ has inherited the scale transformation property from the coupling constant $g \to e^{-c}g$ under the scaling transformation (2.5). Accordingly, for the action $S_{AG}$ to be also invariant under this scale transformation, $A_{\mu\nu\rho\sigma}$ should be also rescaled as

$$G \to e^{-c}G \ , \quad A_{\mu\nu\rho\sigma} \to e^{c}A_{\mu\nu\rho\sigma} \ . \tag{5.2}$$

when other fields and constants are transforming like (2.5) and (4.19) *except for* $g$ now replaced by $G$.

The total 5D bulk action is now $S_{\text{bulk}} \equiv S_2 + S_{AG} \equiv \int d^5x\, (\mathcal{L}_2 + \mathcal{L}_{AG})$. Here $S_2$ is no longer invariant under supersymmetry, but has terms proportional to $\partial_\mu G$, which is supposed to be cancelled by the variation of $S_{AG}$ [1][2]. There are eight sectors contributing to such $\partial_\mu G$-dependent terms out of $\mathcal{L}_2$ after the replacement $g \to G(x)$: (i) gravitino kinetic term, (ii) $(\overline{\psi}_\mu \lambda)\mathcal{D}\varphi$-Noether term, (iii) $(\overline{\psi}_\mu \chi)G_{\nu\rho\sigma}$-Noether term, (iv) $(\overline{\psi}_\mu \psi_\nu)G_{\rho\sigma\tau}$-Noether term, (v) $(\overline{\psi}_\mu \psi_\nu)\widehat{C}$-term, (vi) $(\overline{\psi}_\mu \lambda)\widehat{C}$-term, (vii) $(\overline{\psi}_\mu \chi)\widehat{C}$-term, (viii) $(\overline{\psi}_\mu \psi^{\underline{a}})\widehat{C}$-term. The terms (i) - (iv) contribute, when a derivative $\mathcal{D}_\mu$ hits either some covariant derivatives or field strengths, after a partial integration of the contribution $\delta_Q \psi_\mu \approx \mathcal{D}_\mu \epsilon$, while (v) - (viii) terms contribute, when the derivative hits *hatted* quantities $\widehat{C}^{iJ}$ or $\widehat{\xi}^{\underline{\alpha}}{}_I$. All of these terms containing the derivative $\partial_\mu G$, are therefore cancelled by the appropriate supersymmetry transformation $\delta_Q A_{\mu\nu\rho\sigma}$ in $\delta_Q \mathcal{L}_{AG}$. If we restrict ourselves to the case of linear order supersymmetry transformation of $\delta_Q A_{\mu\nu\rho\sigma}$,[7] then we easily see that only (v) - (viii) terms contribute. For example, (iii) term contributes the cubic combination $\epsilon^{ijk} A_\nu{}^i A_\rho{}^j A_\sigma{}^k$ coming from the Chern-Simons term in $G_{\mu\nu\rho}$, and such terms are omitted from now on. After these considerations, we get

$$\delta_Q\Big(\mathcal{L}_2\Big|_{g\to G}\Big) = \tfrac{1}{24}\epsilon^{\mu\nu\rho\sigma\tau}\Big[\ +2\sqrt{2}e^\sigma(\overline{\epsilon}\gamma_{[\mu\nu\rho|}T^i\psi_{|\sigma]})C^{ij}L^j + \tfrac{i}{\sqrt{2}}e^\sigma(\overline{\epsilon}\gamma_{\mu\nu\rho\sigma}T^i\lambda^a)C^{ij}L_a{}^j$$
$$+ \tfrac{i}{\sqrt{6}}e^\sigma(\overline{\epsilon}\gamma_{\mu\nu\rho\sigma}T^i\chi)C^{ij}L^j + \tfrac{i}{\sqrt{2}}e^\sigma(\overline{\epsilon}\gamma_{\mu\nu\rho\sigma}\psi^{\underline{a}})V_{\underline{\alpha}\underline{a}A}\xi^{\underline{\alpha}i}L^i\Big]\partial_\tau G \ , \tag{5.3}$$

up to cubic or higher order terms. Note that there are *no hats* on $C^{ij}$ and $\xi^{\underline{\alpha}i}$ here, and $L^i$ is the $I=i$ component of $L^I$. Eq. (5.3) is supposed to be cancelled by the new supersymmetry transformation rule $\delta_Q A_{\mu\nu\rho\sigma}$ in $\delta_Q \mathcal{L}_{AG}$:[8]

$$\delta_Q A_{\mu\nu\rho\sigma} = \ -2\sqrt{2}e^\sigma(\overline{\epsilon}\gamma_{[\mu\nu\rho|}T^i\psi_{|\sigma]})C^{ij}L^j - \tfrac{i}{\sqrt{2}}e^\sigma(\overline{\epsilon}\gamma_{\mu\nu\rho\sigma}T^i\lambda^a)C^{ij}L_a{}^j$$
$$- \tfrac{i}{\sqrt{6}}e^\sigma(\overline{\epsilon}\gamma_{\mu\nu\rho\sigma}T^i\chi)C^{ij}L^j - \tfrac{i}{\sqrt{2}}e^\sigma(\overline{\epsilon}\gamma_{\mu\nu\rho\sigma}\psi^{\underline{a}})V_{\underline{\alpha}\underline{a}A}\xi^{\underline{\alpha}i}L^i + \text{(quadratic terms)} \ . \tag{5.4}$$

The fact that the $C$'s and $\xi$'s here have *no hats* is consistent with the scaling property (5.2). As the standard first step of this prescription [2], we require $\delta_Q G = 0$, so that there is no other contribution from $\delta_Q \mathcal{L}_{AG}$. Our previous result [12] can be recovered easily by truncating $\psi^{\underline{a}} \to 0$ and reducing $C^{22} = -1$, $C^{ij}$ (otherwise) $= 0$. Now our action $S_{\text{bulk}}$ is invariant under (5.4), $\delta_Q G = 0$ and (4.17) with $g \to G$.

---

[7]The word 'linear' here does not include the quantities $C^{iJ}$ or $\xi^{\underline{\alpha}}{}_I$. This is because in the reduced case of Abelian $SO(2)$-gauging, $C^{ij}$ is reduced to be a constant, while since $C^{ij}$ has the part $A_{\underline{\alpha}}{}^i \xi^{\underline{\alpha}j}$, and therefore it is more convenient to regard the $\xi$'s as the same order as the $C$'s itself.

[8]The previously-mentioned correction of sign errors in (4.2) in [12] leads to the corrected signs for the first and third terms both in (5.3) and (5.4).

Since we are now dealing with the prescription in [2] originally designed for Abelian gauging without hypermultiplets, applied to our non-Abelian gauging also with hypermultiplets, it is better to confirm the closure of supersymmetry on the field $A_{\mu\nu\rho\sigma}$ by the commutator $[\delta_Q(\epsilon_1), \delta_Q(\epsilon_2)] = \delta_P(\overline{\epsilon}_2 \gamma_m \epsilon_1)$ acting on $A_{\mu\nu\rho\sigma}$, where $\delta_P(\eta_m)$ implies the translation operator. In what follows, we confirm this closure up to quadratic field level. The linear terms in this commutator are composed of six sectors: $F_{\mu\nu}$, $G_{\mu\nu\rho}$, $\partial\sigma$, $\partial\varphi$, $\partial\phi$-linear sectors, and $\widehat{C}^2$ or $\widehat{\xi}^2$-terms. Here, the first four sectors work with no problem, while the $\partial\phi$-linear sector needs special care. To be more specific, we get

$$[\delta_Q(\epsilon_1), \delta_Q(\epsilon_2)] A_{\mu\nu\rho\sigma}\Big|_{\partial\phi} = -e^{-1} e^{\sigma} \epsilon_{\mu\nu\rho\sigma}{}^{\tau} \left[\tfrac{1}{2}(\overline{\epsilon}_1 \epsilon_2) L^i\right] \widehat{\xi}_{\underline{\alpha}}{}^i \mathcal{D}_\tau \phi^{\underline{\alpha}}$$
$$- 2i e^{\sigma} (\overline{\epsilon}_2 \gamma_{[\mu\nu\rho|} T^i \epsilon_1) F_{\underline{\alpha}\underline{\beta}}{}^i \xi^{\underline{\alpha} j} L^j \mathcal{D}_{|\sigma]} \phi^{\underline{\beta}} \ , \tag{5.5}$$

where the last term can be interpreted just as the usual desirable gauge transformation of the type $\partial_{[\mu} \Lambda_{\nu\rho\sigma]}$ up to quadratic terms, while the first term needs special care. This term is actually interpreted as an $Sp(1)$ gauge transformation of $A_{\mu\nu\rho\sigma}$. Even though this seems slightly bizarre at first sight, it can be easily understood, once we notice that the $\phi^{\underline{\alpha}}$-kinetic term is *no longer* $Sp(1)$ invariant after the replacement $g \to G(x)$. In fact, after this replacement, (4.3c) is to be modified as

$$\delta_G(\mathcal{D}_\mu \phi^{\underline{\alpha}}) = \alpha^I (\mathcal{D}_\mu \phi^{\underline{\alpha}}) (\partial_{\underline{\beta}} \widehat{\xi}^{\underline{\alpha}}{}_I) + \alpha^i \xi^{\underline{\alpha} i} \partial_\mu G \ , \tag{5.6}$$

with the new effect of $\partial G$, while all other equations in (4.3) are 'formally' intact. This results in the non-trivial contribution of the $\phi$-kinetic term under the gauge transformation $\delta_G$:

$$\delta_G \left[-\tfrac{1}{2} e g^{\mu\nu} g_{\underline{\alpha}\underline{\beta}} (\mathcal{D}_\mu \phi^{\underline{\alpha}})(\mathcal{D}_\nu \phi^{\underline{\beta}})\right] = \tfrac{1}{24} \epsilon^{\mu\nu\rho\sigma\tau} \left[e^{-1} \epsilon_{\mu\nu\rho\sigma}{}^{\omega} \alpha^i \xi_{\underline{\alpha}}{}^i (\mathcal{D}_\omega \phi^{\underline{\alpha}})\right] \partial_\tau G \ , \tag{5.7}$$

It is now clear that this contribution can be cancelled by an extra transformation $\delta_G A_{\mu\nu\rho\sigma}$ via $\mathcal{L}_{AG}$, such that

$$\delta_G A_{\mu\nu\rho\sigma} = -e^{-1} \epsilon_{\mu\nu\rho\sigma}{}^{\tau} \alpha^i \xi_{\underline{\alpha}}{}^i \mathcal{D}_\tau \phi^{\underline{\alpha}} \ . \tag{5.8}$$

In other words, when we identify $\alpha^i \equiv (1/2)(\overline{\epsilon}_2 \epsilon_1) L^i$ in (5.5), then the first term in (5.5) is absorbed into the $Sp(1)$ gauge transformation.

Even though the result that the tensor potential field $A_{\mu\nu\rho\sigma}$ is transforming under the gauge group $Sp(1)$ seems unnatural at first glance, this is nothing new in supergravity. In fact, in [2] it was pointed out that the original action $S_2$ is no longer $R$-invariant, *i.e.,* in our case $Sp(1)$ *non*-invariant producing a quadratic terms in fermions after the replacement $g \to G(x)$.[9] Analogous situation can be found in Green-Schwarz mechanism in anomaly cancellation in the usual formulation [20][10] or in the dual formulation [21], in which the tensor field $B_{\mu\nu}$ or $M_{\mu_1\cdots\mu_6}$ transforms under Lorentz as well as gauge transformation, as the zero-slope limit effect of superstring theory.

Going back to our closure question, the only left over sector is the $\widehat{C}^2$ and $\widehat{\xi}^2$-terms which turn out to be

$$[\delta_Q(\epsilon_1), \delta_Q(\epsilon_2)] A_{\mu\nu\rho\sigma}\Big|_{\widehat{C}^2} = \tfrac{1}{4}\epsilon_{\mu\nu\rho\sigma\tau} e^{2\sigma} \eta^\tau (C^{ij}\widehat{C}^i{}_K L^{jK} + 2\xi_{\underline\alpha}{}^i \widehat{\xi}^{\underline\alpha}{}_J L^i L^J) \ , \qquad (5.9)$$

where $\eta^\tau \equiv (\overline{\epsilon}_2 \gamma^\tau \epsilon_1)$. If this system is analogous to the Abelian case [2][12], these two terms are supposed to be proportional to $\eta^\tau F_{\tau\mu\nu\rho\sigma}$, upon the use of $G$-field equation, for the field strength of the potential $A_{\mu\rho\sigma\tau}$: $F_{\mu\nu\rho\sigma\tau} \equiv 5\partial_{[\mu} A_{\nu\rho\sigma\tau]}$. In fact, the $G$-field equation is easily obtained as

$$F_{\mu\nu\rho\sigma\tau} = \tfrac{1}{4} e^{-1} e^{2\sigma} \epsilon_{\mu\nu\rho\sigma\tau}(C^{ij}\widehat{C}^i{}_K L^{jK} + 2\xi_{\underline\alpha}{}^i \widehat{\xi}^{\underline\alpha}{}_J L^i L^J) \ , \qquad (5.10)$$

up to quadratic terms. After simple algebra, it is easy to show that (5.9) is desirably proportional to $\eta^\tau F_{\tau\mu\nu\rho\sigma}$ which is equivalent to the combination of the usual translation $\delta_P(\eta^\tau) A_{\mu\nu\rho\sigma}$ accompanied by a gauge transformation. This conclude the linear-order closure of supersymmetry on $A_{\mu\nu\rho\sigma}$, which provides a non-trivial consistency check of our system with $S_{\text{bulk}}$, in particular with non-Abelian gauged groups.

We next consider a possible brane action $S_{\text{brane}}$ to be added. To this end, and for the reason to be clarified later, we truncate the hypermultiplets $(\phi^{\underline{a}}, \psi^{\underline{a}})$, and we restrict the gauged group to be $SO(2)$ out of the $Sp(1)$ isotropy group in the coset $Sp(n',1)/Sp(n') \times Sp(1)$. We do not have to restrict other gauged groups in $G$, but it is only $SO(2)$ out of the $Sp(1)$ group to be gauged.

We next assume that the branes are located at $y \equiv x^5 = 0$ and $y = b > 0$ in the 5-th dimension, requiring all the fields to obey the usual periodic boundary condition $f(-b) = f(0) = f(b)$. Subsequently, we assign the parities under $y \leftrightarrow -y$ on the branes on all the fields in our system, following [1][2]:

$$\Pi(\widetilde{e}_\mu{}^m) = \Pi(e_5{}^{(5)}) = \Pi(A_5{}^I) = \Pi(\sigma) = \Pi(\varphi^\alpha) = \Pi(A_{\mu\nu\rho\sigma}) = \Pi(\psi_\mu) = \Pi(\epsilon) = +1 \ , \quad (5.11\text{a})$$

$$\Pi(e_5{}^m) = \Pi(e_\mu{}^{(5)}) = \Pi(A_\mu{}^I) = \Pi(G) = \Pi(\psi_5) = \Pi(\lambda^a) = \Pi(\chi) = -1 \ , \qquad (5.11\text{b})$$

where $\widetilde{e}_\mu{}^m$ denotes the 4D part of the fünfbein. The parity for an arbitrary bosonic ($\Phi$) or fermionic ($\Psi$) field is defined by [1][2]

$$\Phi(-y) = \Pi(\Phi) \Phi(y) \ , \qquad \Psi(-y) = i\alpha \Pi(\Psi) \gamma_5 T^2 \Psi(y) \ . \qquad (5.12)$$

---
[9]These terms did not matter in our treatment, because we are looking into only linear terms in the transformation rule of $\delta_Q A_{\mu\nu\rho\sigma}$.

Here the real constant $\alpha = \pm 1$ reflects the signature ambiguity, but its sign should be common to all the fermions [1][2].

We now consider the brane action

$$S_{\text{brane}} \equiv -2h \int d^5x \, [\, \delta(y) - \delta(y-b) \,] \Big[ \tfrac{1}{\sqrt{2}} a\widetilde{e} e^\sigma V_I L^I + \tfrac{1}{24} \epsilon^{\mu\nu\rho\sigma 5} A_{\mu\nu\rho\sigma} \Big] \equiv \int d^5x \, \mathcal{L}_{\text{brane}} \, , \quad (5.13)$$

where $h$, $a$ are real constants with $|a|=1$, and $\widetilde{e}$ is the 4D part of the determinant of the fünfbein, while $\epsilon^{\mu\nu\rho\sigma 5}$ is the $\tau=5$ component of $\epsilon^{\mu\nu\rho\sigma\tau}$. The exponential factor $e^\sigma$ is needed for invariance of $S_{\text{brane}}$ under supersymmetry, as will be seen.[10] In order for the action $S_{\text{brane}}$ to be invariant under the scaling transformation (5.2), the constant $h$ should also be rescaled as

$$h \to h^{-c} \, . \quad (5.14)$$

when other fields and constants are transforming like (2.5) and (4.17), *except for $g$ replaced by $G$*.

The action $S_{\text{brane}}$ modifies the $A_{\mu\nu\rho\sigma}$-field equation from the original one $\partial_\mu G = 0$ into

$$\partial_5 G(x) = 2h[\, \delta(y) - \delta(y-b) \,] \, , \quad \partial_\mu G = 0 \quad (\text{for } \mu \neq 5) \, . \quad (5.15)$$

The solution for this field equation is [1][2]

$$G(x) = G(y) = h\,\epsilon(y) = \begin{cases} +h & (\text{for } 0 < y < +b) \, , \\ -h & (\text{for } -b < y < 0) \, . \end{cases} \quad (5.16)$$

This gives the desirable kinks for the $SO(2)$ coupling 'constant' $G(x)$ [1][2].

We now take the variation of $\delta_Q S_{\text{brane}}$ under supersymmetry (4.15) and (5.4) for $\delta_Q A_{\mu\nu\rho\sigma}$:

$$\begin{aligned}
\delta_Q \mathcal{L}_{\text{brane}} = & -2h[\, \delta(y) - \delta(y-b) \,] \tfrac{1}{\sqrt{2}} a\widetilde{e} e^\sigma \big[ (\overline{\epsilon}\gamma^\mu \psi_\mu) - ia^{-1}(\overline{\epsilon}\gamma^\mu\gamma_5 T^2 \psi_\mu) \big] V_I L^I \\
& -2h[\, \delta(y) - \delta(y-b) \,] \tfrac{i}{\sqrt{2}} a\widetilde{e}\epsilon^\sigma \big[ i(\overline{\epsilon}\lambda^a) - a^{-1}(\overline{\epsilon}\gamma_5 T^2 \lambda^a) \big] V_I L_a^{\,I} \\
& -2h[\, \delta(y) - \delta(y-b) \,] \tfrac{i}{\sqrt{6}} a\widetilde{e}\epsilon^\sigma \big[ i(\overline{\epsilon}\chi) - a^{-1}(\overline{\epsilon}\gamma_5 T^2 \chi) \big] V_I L^I \, ,
\end{aligned} \quad (5.17)$$

Comparing these three lines with (5.12), we see that if

$$a = \alpha = \pm 1 \, , \quad (5.18)$$

then (5.17) is simplified to be

$$\begin{aligned}
\delta_Q \mathcal{L}_{\text{brane}} = & -2h[\, \delta(y) - \delta(y-b) \,] \tfrac{1}{\sqrt{2}} a\widetilde{e}(y)\, e^{\sigma(y)} \overline{\epsilon}(y)\gamma^m \big[ \psi_m(y) - \psi_m(-y) \big] V_I L^I(y) \\
& -2h[\, \delta(y) - \delta(y-b) \,] \tfrac{i}{\sqrt{2}} a\widetilde{e}(y)\, e^{\sigma(y)} \overline{\epsilon}(y) \big[ \overline{\lambda}^a(y) - \lambda^a(-y) \big] V_I L_a^{\,I}(y) \\
& -2h[\, \delta(y) - \delta(y-b) \,] \tfrac{i}{\sqrt{6}} a\widetilde{e}(y)\, e^{\sigma(y)} \overline{\epsilon}(y) \big[ \chi(y) - \chi(-y) \big] V_I L^I(y) \, , \quad (5.19)
\end{aligned}$$

---

[10]This factor was not considered in [12].

where each line vanishes after the $\int dy$-integration, under the periodic boundary condition $f(-b) = f(0) = f(b)$ for an arbitrary field $f(y)$ in (5.19). This concludes the proof of the invariance

$$\delta_Q S_{\text{brane}} = 0 \ , \tag{5.20}$$

and therefore that of the total action $S_{\text{total}} \equiv S_2\big|_{g \to G} + S_{AG} + S_{\text{brane}}$ for the $SO(2)$-gauging in the absence of hypermultiplets.

Let us briefly comment on the difficulty of the brane mechanism with the gauging $Sp(1)$ or with hypermultiplets. The difficulty with the $Sp(1)$-gauging is that we do not have a good analog of $V_I L^I$ we can use as the first term in $S_{\text{brane}}$ as an invariant quantity. This is because $\delta_Q A_{\mu\nu\rho\sigma}$ in (5.4) with general $T^i$ matrices with general index $i$ can not cancel the variation $\delta_Q \widetilde{e} = \widetilde{e}(\overline{\epsilon}\gamma^\mu \psi_\mu) + \cdots$ considering the parity (5.12). As for the inclusion of the hypermultiplets, the $\psi^{\underline{a}}$-dependent term in $\delta_Q A_{\mu\nu\rho\sigma}$ yields

$$\delta_Q\left(\tfrac{1}{24}\epsilon^{\mu\nu\rho\sigma\tau} A_{\mu\nu\rho\sigma}\right)\Big|_{\psi^{\underline{a}}} = \tfrac{1}{\sqrt{2}} e^\sigma (\overline{\epsilon}\gamma_5 \psi^{\underline{a}}) V_{\underline{\alpha a} A} \xi^{\underline{\alpha}2} L^2 \ , \tag{5.21}$$

*via* $S_{\text{brane}}$. The trouble here is that there seems to be no invariant lagrangian that will cancel (5.21). For example, a quantity like $\phi^{\underline{\alpha}} \xi_{\underline{\alpha}}{}^2 L^2$ is not appropriate, because of its lack of gauge invariance under (4.3). From these viewpoints, we seem to have no generalization of the brane action $S_{\text{brane}}$ (5.13), when $Sp(1)$ is gauged and/or the hypermultiplets are included.

The brane action $S_{\text{brane}}$ we gave here is supposed to be the simplest one based on [2], among other potentially possible lagrangians invariant under local supersymmetry in singular 5D space-time [1]. However, we stress that our alternative $N = 2$ supergravity in 5D is equally applicable to these formulations, as well.

## 6. Concluding Remarks

In this paper, we have completed the non-Abelian gauging of our alternative $N = 2$ supergravity to $n$ copies of vector multiplets in 5D, and $n'$ copies of hypermultiplets, with a simpler coupling structure compared with the conventional supergravity [6][7][9][8], up to quartic fermion terms in the action. Our result is the combination of considerable works on supergravity couplings in the past, such as vector multiplet couplings in 9D case with the scalars forming the coset $SO(n,1)/SO(n)$ [13][14], together with the scalars in the hypermultiplets forming the quaternionic Kähler manifold in $N = 2$ supergravity in 4D [15] and in 5D [6][7][9][8] as well as in 6D [16].

As in 9D [13], the scalars in the vector multiplets form the coordinates of the $\sigma$-model for the non-Jordan family scalar coset $H^n \equiv SO(n,1)/SO(n)$, and the vector fields with the total number $n+1$ form the $(\mathbf{n+1})$-representation of $SO(n,1)$, while the gaugini $\lambda^a$ form the $\mathbf{n}$-representation of $SO(n)$. The scalars in the hypermultiplets form the $\sigma$-model on the quaternionic Kähler manifold $Sp(n',1)/Sp(n') \times Sp(1)$.

Our result is valid for any arbitrary gauge groups of the type $G \equiv SO(2) \times Sp(n') \times Sp(1) \times \overline{H} \times [U(1)]^{n-p+1}$ for any arbitrary group $\overline{H}$ with $\dim \overline{H} = p - n'(2n'+1) - 4$, and $Sp(n') \times Sp(1)$ are the isotropy groups of the coset $Sp(n',1)/Sp(n') \times Sp(1)$, with a peculiar potential term in the lagrangian. Accordingly, we have obtained a crucial constraint (4.14) required by consistency between the two different cosets under supersymmetry. This constraint relates the $Sp(1)$ curvature $F_{\underline{\alpha}\underline{\beta}}{}^i$ and the $\widehat{C}^{iJ}$-functions, whenever a non-Abelian group in the isotropy groups $Sp(n') \times Sp(1)$ is gauged. Moreover, the isotropy groups $Sp(n') \times Sp(1)$ can be also reduced into their subgroups, e.g., $Sp(1) \times SO(2)$, where $Sp(1) \subset Sp(n')$, $SO(2) \subset Sp(1)$. This a generalization of our previous paper [12], in which we gauged only the $SO(2)$ subgroup of $Sp(1) \subset Sp(n') \times Sp(1)$. Therefore, by adjusting the parameters $n$, $n'$ and $p$ for dimensions appropriately, our results in this paper are considerably general, and cover a wide range of combinations of gauged groups.

Since there are two non-trivial coset structures $SO(n,1)/SO(n)$ and $Sp(n',1)/Sp(n') \times Sp(1)$ present in our system, our vector fields $A_\mu{}^I$ for gaugings are *both* in the $(\mathbf{n+1})$-representation of $SO(n,1)$ *and* in the adjoint representations of the gauged groups in $G$ at the same time. The mutual consistency of these two structures under supersymmetry requires the constraint (4.14) which corresponds to analogous equations in [9][8].

Even though we did not perform explicitly in this paper, we can also combine any Abelian factor groups $[U(1)]^{n-p+1}$, by introducing constant vectors $V_I$, as has been done in [6][7][9][8][12]. This provides another freedom for practical applications for phenomenological model building.

We have also generalized this result to the case of singular 5D space-time for the case of $SO(2)$-gauging as a supersymmetric Randall-Sundrum brane-world scenario [3] similar to the conventional 5D supergravity [1][2]. We have applied the prescription in ref. [2] in order to confirm the supersymmetry of our brane action with the singularity of the type $S^1/\mathbb{Z}_2$. We have also seen some difficulty, when the gauged group is $Sp(1)$ larger than its subgroup $SO(2)$, or when the hypermultiplets are present. As far as the formulation for singular space-time is concerned, there seems to be no fundamental difference between our alternative supergravity and the conventional one [6][7][9][8].

For some readers who wonder why our 'larger' supergravity multiplet [12] has never been studied as a special case in the conventional and 'exhaustively' studied formulation [6][7][9][8], we repeat the following points already given in [12]: The original result in [6] was presented before the discovery of the importance of superstring in 1984 [10], so that there was no strong motivation to include the dilaton or antisymmetric tensor fields, which are important NS fields in terms of superstring language. To put it differently, it is only superstring [10] or M-theory [18] that motivates the peculiar couplings of dilaton and antisymmetric tensor to supergravity, as we have done in the present paper.

Even though we have stressed the difference of our formulation from other general matter couplings in [6][7][9][8], it is fair to point out some similarities. For example, we expect it possible to generalize the number of the additional tensor fields $B_{\mu\nu}$ in addition to the one in the supergravity multiplet with a $\sigma$-model structure similar to that presented in [7][9][8]. However, we also emphasize that our antisymmetric tensor $B_{\mu\nu}$ is still to be distinguished from these additional ones which always appear in pairs [7][9][8]. This situation is analogous to our dilaton $\sigma$ as another NS field separated from other scalar fields.

We believe that our result in this paper will be of great help for the study of the Randall-Sundrum brane-world scenario [3] associated with superstring theory and M-theory.

We are grateful to J. Bagger, A. Ceresole, G. Dall'Agata, S.J. Gates, Jr., M. Günaydin, R. Kallosh, E. Sezgin, P.K. Townsend and M. Zucker for helpful discussions.

## Appendix A: Notations and Conventions

In this Appendix, we deal with important notations and conventions in our paper. Typically, our indices $\mu, \nu, \cdots = 0, 1, \cdots, 4$ are for the curved world indices, and $m, n, \cdots = (0), (1), \cdots, (4)$ are local Lorentz indices with the metric $(\eta_{mn}) = \text{diag.}(-, +, +, +, +)$, and $\epsilon^{012345} = +1$. Most frequently used relevant relations are such as

$$\epsilon_{m_1 \cdots m_{5-n} r_1 \cdots r_n} \epsilon^{r_1 \cdots r_n n_1 \cdots n_{5-n}} = -(n!)[(5-n)!] \delta_{[m_1}{}^{[n_1} \cdots \delta_{m_{5-n}]}{}^{n_{5-n}]} \ , \tag{A.1a}$$

$$\epsilon^{m_1 \cdots m_{5-n} n_1 \cdots n_n} \gamma_{n_1 \cdots n_n} = i(-1)^{n(n-1)/2} (n!) \gamma^{m_1 \cdots m_{5-n}} \ . \tag{A.1b}$$

As the most important notational preparation, we first deal with the fermions in our 5D with the signature $(-, +, +, +, +)$. In this paper, we refer the reader to the general description of arbitrary fermions in diverse space-time dimensions [17]. Using the notation in [17], we start with the number of space and time dimensions together with the parameters $\epsilon$ and $\eta$ that control the properties of fermions [17]:

$$s = 4 \ , \quad t = 1 \ , \quad \epsilon = -1 \ , \quad \eta = -1 \ , \tag{A.2}$$

We next define the complex conjugations by [17]

$$\psi^{\dagger A} = \epsilon^{AB} B \psi_B \ , \quad \psi_B = -B^{-1} \epsilon_{BA} \psi^{\dagger A} \ , \tag{A.3}$$

with the $Sp(1)$ metric $\epsilon^{AB} = -\epsilon^{BA}$ [22][17]. Accordingly [17],

$$\gamma_m^\dagger = -A \gamma_m A^{-1} \ , \quad A \equiv \gamma_{(0)} \ , \quad (\gamma_{(0)})^2 = -I \ , \quad A^\dagger = -AAA^{-1} = -A \ ,$$
$$\gamma_m^\dagger = +\gamma_{(0)} \gamma_m \gamma_{(0)} = +A \gamma_m A \ , \quad \overline{\psi} = \psi^\dagger A = \psi^\dagger \gamma_{(0)} \ , \quad B^\dagger B = I \ ,$$
$$B^T = -B \ , \quad \gamma_m^T = C \gamma_m C^{-1} \ , \quad C^T = -C \ , \quad C = BA \ . \tag{A.4}$$

Relevantly, we have the fermionic quadratic combinations [17]:

$$(\overline{\chi}^A \lambda_A)^\dagger = \lambda_A^\dagger (\overline{\chi}^A)^\dagger = \lambda_A^\dagger (\chi^{\dagger A} A)^\dagger = \lambda_A^\dagger A^\dagger \chi^A = -\lambda_A^\dagger A \chi^A$$
$$= -(\overline{\lambda}_A \chi^A) = -\epsilon^{AB} (\overline{\lambda}_A \chi_B) = -\epsilon^{AB} (\overline{\chi}_B \lambda_A) = -(\overline{\chi}^A \lambda_A) \ . \tag{A.5}$$

Note that we have taken into account a sign error[11] in eq. (6) in [17] that affects the first hand side. In a more general case for $0 \leq n \leq 5$, the hermitian conjugation works like [17]

$$(\overline{\chi}^A \gamma_{m_1 \cdots m_n} \lambda_A)^\dagger = \lambda_A^\dagger (\gamma_{m_1 \cdots m_n})^\dagger (\overline{\chi}^A)^\dagger$$
$$= \lambda_A^\dagger (-A \gamma_{[m_n|} A^{-1})(-A \gamma_{|m_{n-1}|} A^{-1}) \cdots (-A \gamma_{|m_1]} A^{-1})(\chi_A^\dagger A)^\dagger$$
$$= (-1)^{n+n(n-1)/2} (\lambda_A^\dagger A \gamma_{m_1 \cdots m_n} A^{-1} A^\dagger \chi^A) = (-1)^{n+n(n-1)/2} (\overline{\lambda}^A \gamma_{m_1 \cdots m_n} \chi_A)$$
$$= (-1)^{n(n+1)/2} \epsilon^{AB} (\overline{\lambda}_B \gamma_{m_1 \cdots m_n} \chi_A) = (-1)^{n^2/2 + n/2 + n^2/2 - n/2} \epsilon^{AB} (\overline{\chi}_A \gamma_{m_1 \cdots m_n} \lambda_B)$$
$$= (-1)^{n^2 + 1} (\overline{\chi}^A \gamma_{m_1 \cdots m_n} \lambda_A) \ , \tag{A.6}$$

---

[11]We acknowledge E. Sezgin for informing about this error.

Hence, omitting the explicit $A$-indices for contractions as in section 2, we get

$$(\overline{\chi}\gamma_{m_1\cdots m_n}\lambda)^\dagger = (-1)^{n+1}(\overline{\chi}\gamma_{m_1\cdots m_n}\lambda) \ . \tag{A.7}$$

Typical examples are

$$(\overline{\chi}\lambda)^\dagger = -(\overline{\chi}\lambda) \ , \quad (\overline{\chi}\gamma_m\chi)^\dagger = +(\overline{\chi}\gamma_m\chi) \ , \quad (\overline{\chi}\gamma_{mn}\lambda)^\dagger = -(\overline{\chi}\gamma_{mn}\lambda) \ , \quad etc. \tag{A.8}$$

In other words, any combination $(\overline{\chi}\gamma_{m_1\cdots m_n}\lambda)$ with an even number of gammas need a pure imaginary unit 'i' in front to be an hermitian expression, while a combination with an odd number of gammas is already hermitian.

In (A.6), we have used the flipping property for $0 \leq n \leq 5$

$$\begin{aligned}(\overline{\lambda}^A\gamma^{m_1\cdots m_n}\chi_B) &= (-1)^{n+1}(-1)^{(1-n)(2-n)/2}(\overline{\chi}_B\gamma^{m_1\cdots m_n}\lambda^A) \\ &= (-1)^{n(n-1)/2}(\overline{\chi}_B\gamma^{m_1\cdots m_n}\lambda^A) \ . \end{aligned} \tag{A.9}$$

Therefore we have

$$(\overline{\epsilon}_1\gamma^{m_1\cdots m_n}\epsilon_2) \equiv (\overline{\epsilon}_1^A\gamma^{m_1\cdots m_n}\epsilon_{2A}) = \begin{cases} -(\overline{\epsilon}_2^A\gamma^{m_1\cdots m_n}\epsilon_{1A}) & (\text{for } n = 0,\ 1,\ 4,\ 5) \ , \\ +(\overline{\epsilon}_2^A\gamma^{m_1\cdots m_n}\epsilon_{1A}) & (\text{for } n = 2,\ 3) \ . \end{cases} \tag{A.10}$$